\documentstyle[aps,prd,preprint,tighten,floats,epsf]{revtex}

\begin{document}

\preprint{VAND-TH-99-04
\hspace{-30.0mm}\raisebox{-2.4ex}{March 1999}}

\title{Warm Inflation in the Adiabatic Regime - a Model, an Existence
Proof for Inflationary Dynamics in Quantum Field Theory}

\author{Arjun Berera\thanks{E-mail:
ab@ph.ed.ac.uk}
 \thanks{Present address, Department of Physics and
Astronomy, University of Edinburgh, Edinburgh EH9 3JZ, United Kingdom}}

\address{
Department of Physics and Astronomy, Vanderbilt
University,
Nashville, TN 37235, USA}

\maketitle

\begin{abstract}

Warm inflation is examined in a multi-field model. Solutions are
obtained for expansion e-folds and scalar density perturbations.
Nonequilibrium dynamics is restricted to a regime that is displaced only
slightly from thermal equilibrium and in which all macroscopic motion is
adiabatic.  In such a regime, nonequilibrium dynamics is well defined,
provided macroscopic motions that displace the thermal equilibrium state
occur sufficiently slow. The solution has adjustable parameters that permit
observational consistency with respect to expansion e-folds and density
perturbations in the full adiabatic regime, thus insuring a valid solution
regime.  {}For particle physics, 
the model is nonstandard since it requires a large
number of fields, $> 10^4$.  
A particle physics/string interpretation of the model and solutions is
discussed, which can accommodate the large field number requirement.

\vspace{0.34cm}
\noindent
PACS number(s): 98.80 Cq, 05.70.Ln, 11.10.Wx

\end{abstract}

\medskip

hep-ph/9904409

\medskip

In press Nuclear Physics B 2000

\bigskip

\eject

\section{Introduction}
\label{intro}

Inflation is a compelling resolution to the
cosmological puzzles, because it explains how a large class of initial
states evolve into a unique final state that is consistent with our
observed Universe.  The background cosmology that defines inflation is a
Friedmann-Robertson-Walker (FRW) cosmology in which the scale factor $R(t)$
has positive acceleration ${\ddot R}(t) > 0$.  The importance of such a
background cosmology for one of the cosmological puzzles, the horizon
problem, has been appreciated for the longest 
time \cite{horizon,oldphaset} (for more details please see
\cite{olive}). 
Later, Guth in his
pivotal paper \cite{guth} pointed out the importance of this background
cosmology, which he named inflation, to solving another cosmological
puzzle, the flatness problem.  Moreover, Guth's paper drew attention to
the relevance of cosmological phase transitions in attaining a dynamical
particle physics explanation for inflation.  These ideas were
foreshadowed by some earlier works \cite{oldphaset}.  
{}Furthermore, most of these
ideas about cosmological phase transitions as well as 
suggestions about their importance in
explaining the cosmological puzzles were initially developed
by Kirzhnits and Linde \cite{kl}.  The thermal field theory and its
application to cosmological phase transitions was considered further by
Weinberg \cite{weinberg} and Dolan and Jackiw \cite{dj}.

These early ideas all converged into a single benchmark concept about
inflationary dynamics, the slow-roll scalar field scenario \cite{ni,ci}.
This picture postulates the existence of a scalar field
$\phi({\bf x},t)$, named the
inflaton, which governs inflationary dynamics.  At some early time, it is
assumed that the energy density $\rho$ and pressure density p were
dominated by the homogeneous component of the inflaton $\varphi_0(t)$, 
where, in
addition, $\varphi_0(t)$ is assumed classical, with
\begin{equation}
\rho = V(\varphi_0) + \frac{1}{2} {\dot \varphi}_0^2 + \rho_r
\label{rho}
\end{equation}
and
\begin{equation}
p = -V(\varphi_0) + \frac{1}{2} {\dot \varphi}_0^2 + \frac{1}{3} \rho_r.
\label{p}
\end{equation}
$V(\varphi_0)$ and ${\dot \varphi}_0^2/2$ are respectively 
the potential and kinetic energy of the inflaton and $\rho_r$ is a
component of radiation energy density.  The key observation for
inflationary dynamics is when the potential energy dominates,
$V(\varphi_0) > {\dot \varphi}_0^2/2, \rho_r$, the equation of state
from Eq. (\ref{rho}) and (\ref{p}) becomes that of the vacuum
$\rho \approx -p$, and this is the required condition in FRW cosmology
for inflationary expansion.  In the slow-roll inflationary scenario, 
considerations from observation generally require the  
stronger condition $V(\varphi_0) \gg {\dot \varphi}_0^2/2$.
This condition along with the assumption that the potential is slowly
varying implies the
equation of motion for $\varphi_0$ is first order in time.

The conditions on $\rho_r$ lead to two different types of thermodynamic
regimes of inflation.  {}For $\rho_r \approx 0$
expansion is isentropic during inflation,
so that the Universe rapidly becomes supercooled. 
In this supercooled inflation regime, the termination of the
inflationary period into a radiation dominated period occurs through a
short intermediate reheating period, in which all the vacuum energy of
the inflaton is converted into radiation energy.
Alternatively, for $V(\varphi_0) \equiv \rho_v > \rho_r >0$, expansion is
non-isentropic during inflation, so that the temperature of the Universe
may still be sizable.  In this warm inflation regime,
conversion of vacuum energy into radiation energy occurs throughout the
inflationary period and the inflationary regime smoothly terminates into
a radiation dominated regime without an intermediate reheating
period.  Warm inflation cosmology, i.e., inflationary dynamics without
reheating, was formulated in \cite{wi,ab1}.  An earlier paper by Fang and the
author \cite{bf2} developed some underlying ideas, although the
statement of that paper is general to all of inflationary cosmology.
The demonstration that non-isentropic expansion, the background
cosmology of warm inflation, can be realized in FRW cosmology from a
plausible field theoretic dynamics was given in \cite{ab2,rudnei,gmn}.

Although the scale factor dynamics of the background cosmology in both
supercooled and warm inflation regimes is similar, the microscopic scalar
field dynamics in the two regimes is very different.
In warm inflation dynamics, during the
inflation period the inflaton interacts considerably with
other fields.  These interactions permit energy exchange and this is how
the vacuum energy of the inflaton is converted into radiation energy.
As the fields acquire the vacuum energy liberated by the inflaton
and become excited, their reaction back on the inflaton damps its
motion.  To realize a viable inflationary regime, the inflaton must
support the vacuum energy sufficiently long to solve the
horizon/flatness problems, $N_e \stackrel{>}{\sim} 60$. 
In warm inflation this implies
the reaction of all the fields on the inflaton must be sufficiently
strong to overdamp its motion.  Thus slow-roll motion in warm inflation
is synonymous with overdamped motion.  {}Furthermore,
the effective dynamics of the
inflaton is analogous to time dependent Ginzburg-Landau scalar
order parameter kinetics.  Such a kinetics has been derived from a
near-thermal-equilibrium quantum field theory
formulation \cite{bgr}.  In this derivation, the classical inflaton is
defined as the thermal expectation value of the inflaton field operator
$\phi({\bf x},t)$, 
$\varphi({\bf x},t) \equiv <\phi({\bf x},t)>_{T}$.

In supercooled inflation dynamics, the inflaton typically is modeled as
noninteractive with other fields during the inflation period.
Supercooling implies the Universe is in the ground state $|0>$ during
inflation. It is assumed that the vacuum is translationally invariant.
The classical inflaton that emerges in the slow-roll equation and
energy-pressure densities, Eqs. (\ref{rho}) and (\ref{p}), is defined as
the expectation value of the inflaton field operator with respect to 
$|0>$, $\varphi({\bf x},t) \equiv <0|\phi({\bf x},t)|0>$.
Several authors \cite{guthpi2,lyth,halliwell} 
have argued on the basis of saturating the
momentum-position uncertainty principle that the quantum equations
for $\phi$ go over into classical equations for a given inflaton
mode of comoving momentum ${\bf k}_c$, $\phi_{{\bf k}_c}(t)$,
once the physical momentum associated with the mode
${\bf k}_p \equiv {\bf k}_c/R(t)$ crosses the Hubble radius
$2\pi/{\bf k}_p > 1/H$.

The slow-roll scalar field scenario has a second aspect to it.
Small fluctuations of the inflaton about its homogeneous component
provide the initial seeds of density perturbations \cite{guthpi1,amp}.
These density perturbations produced during inflation evolve into the
classical inhomogeneities observed in the Universe.  {}For warm inflation
dynamics, the fluctuations of the inflaton are thermally induced. As
such, these initial seeds of density perturbations already are classical
upon inception.  In supercooled inflation dynamics, the fluctuations of
the inflaton arise from zero-point quantum fluctuations and so are
purely quantum mechanical.  In this case, it must be explained how these
initial quantum fluctuations evolve into the classical inhomogeneities
observed in the Universe.  This problem often is referred to as the
quantum-to-classical transition problem of supercooled inflation.  
Early resolutions to this
problem followed the same reasoning as for the homogeneous component.
Based on uncertainty principle arguments for the field amplitude and its
conjugate momenta, these arguments concluded that the fluctuations can
be treated classically once the physical wavelength of a give mode is
larger than the Hubble radius \cite{guthpi2,lyth,halliwell}.
Sasaki noted \cite{sasaki} that this criteria for classical behavior is not
invariant under canonical transformations.  Thus unless a definite
physical significance could be ascribed to the field amplitude and its
conjugate momenta, such arguments are insufficient for explaining the
classical realization of the quantum process.

The complete resolution to the quantum-to-classical problem 
in supercooled inflationary dynamics has been
understood to occur via a process which eliminates quantum interference
between macroscopically distinguishable events. Such a process
generally is termed decoherence \cite{decoher}. 
A common way to
introduce decoherence is through an external environment with which the
inflaton interacts.  Various studies have examined decoherence along
these lines in which external fields couple to the inflaton
\cite{branddec} and even in which the short wavelength modes of the
inflaton act to decohere the long wavelength modes \cite{nambu}.

{}For warm inflation, since dissipation is a fundamental aspect of the
dynamics, decoherence is an automatic consequence.  So warm inflation
contains a good example of decoherence within an inflationary dynamics.
{}For supercooled
inflation, decoherence is not a natural requirement of the dynamics, but
rather it must be imposed an an additional condition.  Thus to the extent of 
relevance of decoherence,
warm inflation realizes the full consequences of it both to yield a
classical dynamics and to exploit the fluctuation-dissipation effects
associated with it.

In the basic warm inflation picture, the only requirement is that
radiation energy is produced during inflation and that the mechanism of
production is via dissipative effects on the inflaton.  
Up to now, the only quantum
field theory realization of this picture is when the radiation energy is
near thermal equilibrium \cite{bgr,bgr2}.  
This is an interesting regime for developing
a warm inflation dynamics, since it is the best understood regime of
nonequilibrium quantum field theory.  {}For the most part, there is
limited understanding of nonequilibrium quantum field theory. However
near thermal equilibrium, the advantage is that the state of the
system can be
studied as a perturbation about the thermal equilibrium state.  
{}Furthermore, if all macroscopic motion is slow relative to the relevant
microscopic time scales, the macroscopic dynamics can be treated
adiabatically.  {}Finally, if the interactions are weak, perturbation
theory is applicable.  In such a thermalized, adiabatic, perturbative
regime, a well formulated quantum field theory dissipative dynamics can
be formulated.  The foundations for this were developed by Kubo
\cite{kubo} and Zubarev \cite{zubarev}, who basically examined the full
dynamical consequences of the near-thermal-equilibrium
fluctuation-dissipation theorem
\cite{fdt}.  Quantum mechanical models that implement the 
fluctuation-dissipation theorem within a dissipative dynamics have
been known for a long time \cite{mag}, although in recent times such
models generically have been
termed Caldeira-Leggett models \cite{cl,kac}. They have been
studied in relativistic models by various authors \cite{clstudies}.

Within a realistic scalar quantum field theory model, Hosoya and
Sagagami \cite{hs1} initially formulated dissipative dynamics. In their
formulation, the near-thermal-equilibrium dynamics is expressed through
an expansion involving equilibrium correlation functions.  Although their
formulation is physically transparent, formally it is cumbersome.
Subsequently Morikawa \cite{morikawa} formulated the same problem in
terms of an elegant real time finite temperature field theory formalism.
These works were developed further by Lawrie \cite{lawrie} and
Gleiser and Ramos \cite{gr}.
{}For warm inflation, the overdamped regime of the inflaton is
the one of interest.
A realization of overdamped motion within a quantum field theory model 
based on this formulation was obtained in \cite{bgr}.

An application of a near-thermal-equilibrium fluctuation-dissipation
dynamics for
warm inflation first was examined from a Caldeira-Leggett type model
\cite{ab1}.  {}For a realistic quantum field theory model,
the results of \cite{bgr} were applied to warm inflation in \cite{bgr2}
and a solution to the horizon/flatness problem was presented.

The warm inflation solution in \cite{bgr2} was based on a specific
quantum field theory model that generally has been termed distributed
mass models (DM-models) \cite{bgr,abpas}.
In this model, the inflaton interacts with several other 
fields through shifted couplings $g^2(\varphi_0-M_i)^2 \chi_i^2$ and
$g(\varphi_0-M_i) {\bar \psi}_i \psi_i$ to bosons and fermions repectively.
The mass sites $M_i$ are distributed over some range.
As the inflaton relaxes toward its minimum energy configuration, it decays
into all fields that are light and coupled to it.  In turn this generates 
an effective viscosity.
In order to satisfy the e-fold requirement of a successful inflation,
$N_e > 60$,
overdamping must be very efficient. The purpose
of distributing the masses $M_i$ is to increase the interval
for $\varphi_0$ in which light particles emerge through the shifted
couplings.  

Within this simple near-thermal-equilibrium quantum field theory
formulation, the distribution of mass sites appears necessary to sustain the
inflaton's overdamped motion sufficiently long to satisfy the e-fold
requirement.  On the one hand, the DM-model may be regarded as 
an intermediate step towards realistic warm inflation models. On the
other hand, the hierarchy of mass sites $M_i$ in this model is
suggestive of mass levels of a fundamental string. Based on this
hypothesis, it was shown in \cite{bk} that DM-models can be obtained
from effective supersymmetric theories.  {}Further development of the string
interpretation of DM-models is given in \cite{bk} and \cite{bk2}.
In this paper we study a wide range of warm inflation solutions
for the DM-models, that extends the single case studied in
\cite{bgr2}.  The specifics of the extensions are clarified in the
sections to follow.  In addition, here the first estimates are given
for thermally induced density perturbations \cite{bf2} in the DM-models.

There is a second aspect to the present work, which perhaps is more
important.  It was mentioned above that the near-thermal-equilibrium
quantum field theory formalism developed in
\cite{hs1,morikawa,lawrie,gr} is unambiguously valid within a
thermalized, adiabatic, perturbative regime.  However in the basic
formulation, the criteria for consistency are specified in terms of
only a set of limiting inequalities (i.e., $\ll$, $\gg$).
As such, these criteria are not specific
about the extent to which the inequalities must be satisfied.
In light of this, a convincing proof that a given dynamics is
consistent with this formalism is if the solution space of interest
exists for an arbitrary degree of validity for the consistency
inequalities.  {}For warm inflation dynamics, the solution space of
interest is the regime of observational consistency with respect to
expansion e-folds, $N_e \stackrel{>}{\sim} 60$, and density perturbations,
$\delta \rho/\rho \stackrel{<}{\sim} 10^{-5}$ 
\cite{smoot} (for a review of the basic
observational facts please see \cite{brandrev,olive,kolb}).
In this paper, we show that the consistency inequalities of the
underlying formalism are satisfied to an arbitrary degree 
and alongside this, an observationally consistent warm inflation regime
always exists.  In particular, as will be seen, 
the most restrictive consistency
conditions involve the adiabaticity requirements. In the solutions, it
will emerge that the degree of adiabaticity can be controlled by one
parameter, $\alpha$, with adiabaticity improving as $\alpha \rightarrow
0$. What will be shown is in this limit, an observationally consistent
warm inflation regime with respect to $N_e$ and $\delta \rho/\rho$
always exists.  This result has a fundamental relevance, since it is an
existence proof within quantum field theory for 
observationally consistent inflationary dynamics from start to finish: from
an initial radiation dominated or inflationary regime, to an inflationary
regime and finally into a radiation dominated regime.

The paper is organized as follows.  In Sect. \ref{sect2} the DM model
Lagrangian is presented and useful results about its effective potential
are reviewed.  In Sect. \ref{sect3} the basic equations of warm
inflation and the consistency conditions on the solutions are reviewed.
Subsect. \ref{subsect3C} provides a convenient summary of all parameters,
notation, and terminology used in this paper.  In Sects.
\ref{sect4} and \ref{sect5} the solutions for e-folds and density
perturbations respectively are presented.  Subsects. \ref{subsect4C} and
\ref{subsect5C} discuss general features about the solutions in order to
ease the effort to understand the calculations.  This paper will not
focus on phenomenological consequences of the model.
In Sect. \ref{sect6} some example applications of the solutions are given.
Subsect \ref{subsect6A} computes the dimensional scales of the relevant
quantities in the cosmology.  Subsect. \ref{subsect6B} analyzes this
warm inflation model in the limit of arbitrary adiabaticity. Subsect.
\ref{subsect6C} examines the solution's dependence on $\lambda$, the $\phi$
self coupling parameter.
In the concluding section \ref{sect7}, first
improvements to this calculation
are discussed. Second, a  particle/string physics interpretation of the
model is discussed.  {}Finally some concluding  perspectives are given
about the model and solutions.

\section{Model}
\label{sect2}

Consider the following Lagrangian of a scalar field $\phi$
interacting with $N_M\times  N_{\chi}$ scalar fields $\chi_{ik}$ and
$N_M\times  N_{\psi}$ fermion fields $\psi_{ik}$,
\begin{eqnarray}
{\cal L} [ \phi, \chi_{ik}, \bar{\psi}_{ik}, \psi_{ik}, \chi^r_{i},
\bar{\psi}^r_{i}, \psi^r_{i} ] & = &
\frac{1}{2}
(\partial_\mu \phi)^2 - \frac{m^2}{2}\phi^2 -
\frac{\lambda}{4 !} \phi^4 - V_0 - V_1(\phi) \nonumber\\
& + & \sum_{i=i_{\rm min}}^{i_{\rm max}} \sum_{k=1}^{N_{\chi}} \left\{
\frac{1}{2} (\partial_\mu \chi_{ik})^2
- \frac{f_{ik}}{4!} \chi_{ik}^4 - \frac{g_{ik}^2}{2}
\left(\phi-M_i\right)^2
\chi_{ik}^2 
\right\}
\nonumber \\
& + & \sum_{i=i_{\rm min}}^{i_{\rm max}} \sum_{k=1}^{N_{\psi}}  
\left\{ i \bar{\psi}_{ik} \not\!\partial \psi_{ik} - h_{ik} ( \phi - M_i ) 
\bar{\psi}_{ik} \psi_{ik} \right\} \nonumber\\
& + & \sum_{i=1}^{N_r} \left\{
\frac{1}{2} (\partial_\mu \chi^r_i)^2
- \frac{f^r_{i}}{4!} {\chi^r_{i}}^4 \right\} +
\sum_{i=1}^{N_r/4}   
\left\{ i \bar{\psi}^r_{i} \not\!\partial \psi^r_{i} \right\}  
\: .
\label{Nfields}
\end{eqnarray}
\noindent 
This model is described in \cite{bgr,abpas}
and named the distributed mass
model (DM model), since the interaction between $\phi$ with the $\chi_{ik}$
and $\psi_{ik}$
fields establishes a mass scale distribution for the $\chi_{ik}$ and
$\psi_{ik}$ fields,
which is determined by the mass parameters $\{M_i\}$. 

The finite temperature effective potential for this model was computed in
\cite{bgr,bgr2} for a nonzero homogeneous field amplitude
$\langle \phi ({\bf x},t) \rangle_T \equiv \varphi_0(t)$.
Because of the shifted coupling arrangement,
the self-coupling parameter $\lambda$ is not corrected by $\chi - \phi$
and $\psi-\phi$ interactions.
However, these interactions will generate 
new terms $\propto (\phi-M_i)^n$ and
they are discussed in the next section.  The finite temperature, field
dependent mass are
\begin{equation}
m_{\chi_{ik}}^2 (\varphi_0,T) = g^2(\varphi_0-M_i)^2 + 
\mu_{\chi_{ki}}^2(T),
\label{rmchi}
\end{equation}
\begin{equation}
m_{\psi_{ik}}^2 (\varphi_0,T) = [g(\varphi_0-M_i) + \mu_{\psi_{ki}}(T)]^2
\label{rmpsi}
\end{equation}
and
\begin{equation}
m_{\phi}^2(\varphi_0,T) = m^2 +\frac{\lambda \varphi_0^2}{2}
+\mu_{\phi}^2(T).
\label{rmphi}
\end{equation}
$\mu(T)$ are thermal mass corrections 
with $\mu_{\phi}(T) \sim \sqrt{\lambda}T$
and $\mu_{\chi,\psi}(T) \sim gT$.  These thermal masses are nonzero only
when the field dependent contribution to the respective particle's mass is
below the temperature scale.

Note that the $\chi_{ik}$
and $\psi_{ik}$ effective field-dependent masses, 
$m_{\chi_{ik},\psi_{ik}} (\varphi_0,T)$,
can be constrained
even when $\langle \phi \rangle = \varphi_0$ is large.
The $\phi\chi$, $\phi\psi$ 
interactions can be
made reflection symmetric, $\phi \rightarrow -\phi$, but for our purposes
we consider all $M_i >0$ and 
$\varphi_0>0$.  The parameters $M_i$ will
be referred to as mass sites.  The $\chi$ and $\psi$ fields will be
referred to as dissipative heat bath fields.
 
{}For our calculations we will consider all coupling constants 
to be positive: $\lambda$,
$f_{ik},g_{ik}^2, h_{ik}, f^r_i$ $> 0$, and 
for simplicity choose them to be
$f^r_i=f_{ik}=f$, $g_{ik}=h_{ik}=g$.
Also, we will set $N_{\chi}=N$, $N_{\psi}=N/4$, which implies an equal number of
bose and fermi degrees of freedom at each mass site $M_i$.  This relation on the
degrees of freedom
along with our choice of coupling implies a cancellation
of radiatively generated vacuum energy corrections in the effective
potential \cite{bgr2}.
{}For convenience we will choose the mass scales to
be evenly spaced as $M_i = i M/g$, 
$i=i_{\rm min}, \ldots, i_{\rm max} \equiv i_{\rm min} + N_M$, where
$M$ is the mass splitting scale between adjacent sites.
Here $M_{i_{\rm min}}$ and $M_{i_{\rm max}}$ bound the interval for
$\varphi_0$ in which dissipative dynamics, thus warm inflation,
is possible.

The Lagrangian Eq. (\ref{Nfields}) also contains $N_r$ bosonic $\chi^r$ and
$N_r/4$ fermionic $\psi^r$ fields that do not interact with the
inflaton $\phi$. These fields represent 
additional degrees of freedom in the heat
bath that otherwise do not contribute to the dissipative dynamics of $\phi$.
These fields will be referred to as non-dissipative heat bath fields.  Our
choice of equal numbers of bose and fermi degrees of freedom 
for the non-dissipative heat bath fields 
is not necessary, but done here for notational
convenience.  

The non-dissipative heat bath fields as written in the above 
Lagrangian are completely noninteracting with either
$\phi$ and the dissipative heat bath fields 
coupled to $\phi$. Thus, in the form written,
none of the energy liberated by the inflaton $\phi$ is transferred into the
non-dissipative heat bath fields.  
To properly realize such heat bath fields, decay channels are
necessary between them and the system of fields coupled to $\phi$. 
We will not attempt to correct this shortcoming but assume that it is
possible. Up to the level of approximation in \cite{gr,bgr,bgr2}, which
is what is applied here, this problem can be solved.  Here,
simply we want to explore the consequences of this
possibility. As will be seen, observationally consistent warm inflations 
can be achieved with no non-dissipative heat bath fields $N_r=0$, and
in many cases this is the optimal situation. However, as also will be seen, 
interesting possibilities arise when there are a very large number 
of non-dissipative heat bath
fields.

The Lagrangian Eq. (\ref{Nfields}) has two extensions to the one
in \cite{bgr2}.  {}First the above Lagrangian allows for an
overall constant shift in the vacuum energy $V_0$. Second,
an additional term has been added to the interaction potential,
$V_1(\phi)$.  This additional term is defined to be zero
within the interval of $\varphi_0$ in which we study warm inflation,
$M_{i_{\rm min}} < \varphi_0 < M_{i_{\rm max}}$.
However outside this region, $V_1(\phi)$ will be suitable to permit
an absolute minimum of the $\varphi$-efffective potential at zero potential
energy.  In the calculations to follow, one case is treated where $V_0$ and
$V_1(\phi)$ are nonexistent, $V_0=0$ and $V_1(\phi)=0$ for all $\phi$.

A few comments are in order to motivate the extensions $V_0$ and
$V_1(\phi)$.  The focus of this study is to understand the quantum field
theory dynamics that underlies warm inflation.  In this respect, the
extensions $V_0$ and $V_1(\phi)$, permit exploration of warm inflation
dynamics in a variety of interesting regimes.  In this paper, we will
not attempt to motivate these extensions from particle physics.
However, note that the effect of $V_0$ and $V_1(\phi)$ is similar to a
plateau region that could arise from higher polynomial or non-polynomial
potentials.  Non-polynomial potentials can occur in particle physics
from nonperturbative effects. {}For example, nonperturbative mechanisms
for dynamical SUSY breaking involving instantons can yield
non-polynomial potentials (for a review please see \cite{shifvan}).
{}Finally note that the basic results in this paper do not rely on these
extensions $V_0$ and $V_1(\phi)$. A warm inflation regime consistent
with quantum field theory and observational requirements on e-folds and
density perturbations can be obtained without $V_0$ and $V_1(\phi)$,
i.e. $V_0=V_1(\phi) =0$. This case is treated in Subsects.
\ref{subsubsect4B2} and \ref{subsubsect5B2}.

\section{Statement of the Problem}
\label{sect3}
    
Let $t=t_{BI}=0$ define the time when warm inflation begins ($BI$) and
$t=t_{EI}$ define the time when warm inflation ends ($EI$).
The thermodynamic state of the Universe during warm inflation is given
by the time dependent temperature $T(t)$. The results presented in
the next sections require two temperature scales $T_{BI} \equiv
T(t_{BI}=0)$
and $T_{EI} \equiv T(t_{EI})$, 
the temperatures of the universe at the beginning
and end of warm inflation respectively.  These two
temperatures will be expressed through the ratios
\begin{equation} 
\beta \equiv \frac{T_{EI}}{T_{BI}}
\label{betadef}
\end{equation}
and 
\begin{equation} 
\kappa_M \equiv \frac{T_{BI}}{M},
\label{kapdef}
\end{equation}
where $M$, defined in Sect. \ref{sect2}, is the splitting scale 
between adjacent mass sites in the DM model.

\subsection{Basic Equations}
\label{subsect3A}

The basic equation for warm inflation dynamics
is an effective equation of motion for
the scalar field $\varphi_0$, which in the warm inflation regime
describes overdamped motion.   This equation has been derived for
DM-models in \cite{bgr2} with details about the bosonic disspative term in
\cite{bgr} and the fermionic dissipative term in \cite{yl}.  Intuitive
explanations of the formalism are given in \cite{hs1,yl}.
In our calculations,
the contribution from fermionic dissipation is ignored.  The only 
effect considered from
fermions is their contributions to the temperature
dependent $\varphi$-effective potential. The most important term 
they contribute is their $T=0$
radiatively generated vacuum energy term, 
since, due to our choice of bose-fermi degrees of freedom,
it cancels a similar term from the bose sector.   The high-T modifications
from the fermion fields are not essential for the overall consistency of the
calculation and they also do not cause any new problems.  We will account for
these contributions in the $\varphi$-effective potential.  
Had we accounted for the fermion effects to dissipation, 
the damping of $\varphi_0$ would increase, which in turn would 
increase the robustness of warm
inflation solutions.  However in our case, where there are one-forth the number
of fermion versus bosonic fields, for the most part,
based on the high-T expressions for the
dissipative terms \cite{bgr,bgr2,yl}, their contribution would
enhance disspation by only 20 percent.  We believe one can configure situations
in which the fermion versus bose contribution to dissipation dominates, which is
interesting since the functional behavior 
of the two types of dissipation differ.
This is another possibility to examine, especially since the analysis in
\cite{yl} found greater success with fermionic dissipation.  
However here we will
not pursue that direction.

{}For all the cases studied below,
the basic $\varphi_0$
effective equation of motion
is from \cite{bgr,bgr2}
\begin{equation}
\sum_i^{t.e.}(\varphi_0-M_i)^2 \eta_{1i}^B(T)
\frac{d \varphi_0}{dt} = 
- \frac{\partial V(\varphi_0,T)}{\partial \varphi_0},
\label{eom1}
\end{equation}
where $t.e.$ means sum over all thermally excited sites.  
Everything multiplying $d\varphi_0/dt$ will be
referred to as the dissipative coefficient and $\eta_{1}^B(T)$ will be
referred to as the dissipative function.
Here $V(\varphi_0,T)$ is the effective potential for $\varphi_0$.
Both the dissipative coefficient and $\varphi$-effective potential emerge
upon integrating out the dissipative heat bath fields $\chi_{ik}$ and
$\psi_{ik}$.
For $\chi_{ik}$ - fields with $m_{\chi_i} < T$, the high-T
expression for $\eta_{1i}^B$ is
\begin{equation}
\eta_{1i}^{B}(T) \stackrel{T \gg m_{\phi},m_{\chi_{ik}} }{\approx} 
\frac{\eta'Ng^4}{\pi T {\cal C}(g,f)},
\label{eta1B}
\end{equation}
where
\begin{equation}
{\cal C}(g,f) \equiv g^4\left[1-\frac{6}{\pi^2}{\rm Li}_2
\left(\frac{m_{\chi}(\varphi_0,T)}{m_{\phi}(\varphi_0,T)}\right)
\right]+ \frac{f^2}{8} \approx g^4+\frac{f^2}{8}
\label{calc}
\end{equation}
and $\eta'= 48 \ln(2T/\mu_{\chi})$.
In an expanding background, the interaction of $\phi$ with the metric
also yields a $3H{\dot \varphi}_0$ term in the $\varphi_0$-effective
equation of motion.  As discussed in \cite{bgr,bgr2}, this term is
dropped in Eq. (\ref{eom1}) because the thermal dissipation term
dominates.

In order to be within an inflationary regime, $V(\varphi_0,T)$ must be
vacuum energy dominated, so that 
$V(\varphi_0,T) \approx \rho_v(\varphi_0)$, where $\rho_v(\varphi_0)$
is the vacuum energy of the $\phi$-field,
$\rho_v(\varphi_0) = V_0+\rho_{\phi}(\varphi_0)$,
where $V_0$ is the vacuum energy density shift parameter
and $\rho_{\phi}(\varphi_0) = V(\varphi_0,T=0)$.
In this case, the rate
of decay of vacuum energy is obtained as 
${d \rho}_v/{dt} = V'(\varphi_0) {\dot \varphi_0}$. The energy
balance equation describing the transfer of vacuum energy to
radiation energy is
\begin{equation}
{\dot \rho}_r = -4H \rho_r - {\dot \rho}_v.
\label{dotrex}
\end{equation}
In the warm inflation regime $\rho_r$ decreases slowly,
so that to a good approximation Eq. (\ref{dotrex}) becomes
\begin{equation}
4H \rho_r \approx - {\dot \rho}_v.
\label{dotr}
\end{equation}
{}For the four cases considered in Sects. \ref{sect4} and \ref{sect5}, 
we examine the regime
where the $\lambda \varphi_0^4$ term dominates the effective
potential. 

\subsection{Consistency Conditions}
\label{subsect3B}

The validity of the basic equations of warm inflation given above
requires four consistency checks, which we refer to
as the thermalization, adiabatic, force and infra-red conditions.
Below, these conditions are explained and exactly specified.

The thermalization and adiabatic conditions both address the consistency
of the macroscopic warm inflation equations with the underlying
microscopic quantum field theory dynamics.
Within the simple 
dissipative quantum field theory
formulation in \cite{bgr}, the decay widths $\Gamma_{\phi}$,
$\Gamma_{\chi}$, and $\Gamma_{\psi}$ characterize the time
scale of microscopic dynamics.  The basic self-consistency
requirement is that all microscopic time scales are faster than
all macroscopic time scales.  Thus the slowest microscopic
time scale is the important one for checking self-consistency.
{}For dissipative dynamics, $\Gamma_{\phi}$ does not
play a significant role. 
Since we are restricting to the bosonic contribution to dissipation
\footnote{Since the fermion contribution to dissipation is
being ignored,  the behavior of $\Gamma_{\psi}$ is not very essential.
If $\Gamma_{\psi} < \Gamma_{\chi}$, then the $\psi$-fields may not be thermally
excited, which in turn would slightly modify the temperature dependent terms in
the effective potential.  However since the $\psi$-fields are free from the
crucial self-consistency requirements for producing dissipative effects on
$\varphi_0$, if $\Gamma_{\psi} < \Gamma_{\chi}$, is is easy to increase 
their number of decay channels, which thereby can increase 
$\Gamma_{\psi}$ to the level of $\Gamma_{\chi}$.},
$\Gamma_{\chi}$ is the relevant microscopic rate.
{}From \cite{bgr}, the high temperature expression for it is
\begin{equation}
\Gamma_{\chi}(T) \stackrel{T \gg m_{\phi}, m_{\chi} }{\approx} 
\frac{\Gamma' T}{\pi} {\cal C}(g,f),
\label{chidecay}
\end{equation}
where ${\cal C}$ is defined in Eq. (\ref{calc})
and $\Gamma' = 1/192$.  {}For $\chi$-fields at site $i$, their decay rate is
Eq. (\ref{chidecay}) when 
$m_{\chi_{ik}} \stackrel{<}{\sim} T$ and negligible when
$m_{\chi_{ik}} \stackrel{>}{\sim} T$. 

Hereafter, rescaled time is used
\begin{equation}
\tau \equiv \Gamma_{\chi}(M) t.
\label{ttau}  
\end{equation}
The Hubble parameter at the beginning of warm inflation will be
expressed as
\begin{equation}
H_{BI} = \alpha \beta \Gamma_{\chi}(T_{BI})
= \alpha \beta \kappa_M \Gamma_{\chi}(M),
\label{hubdef}
\end{equation}
where the additional parameter $\alpha$ is introduced and
$\beta$ and $\kappa_M$ are defined in Eqs. (\ref{betadef}) and
(\ref{kapdef}) respectively.

Warm inflation fundamentally has two macroscopic time scales,
the Hubble expansion rate and the rate of change of the
inflaton, which expressed in terms of rescaled time Eq. (\ref{ttau})  
is ${\dot \varphi}_0 = \Gamma_{\chi}(d\varphi_0/d\tau)$. 
Self consistency requires the
{\bf thermalization condition}
\begin{equation}
H(\tau) \ll \Gamma_{\chi}(T)
\label{therm}
\end{equation}
and the $\varphi_0$-{\bf adiabatic condition}
\begin{equation}
\frac{1}{\varphi_0} \frac{d \varphi_0}{d\tau} \ll 
\frac{\Gamma_{\chi}(T)}{\Gamma_{\chi}(M)}.
\label{padiab}
\end{equation}
{}For the cases in Sects. \ref{sect4} and \ref{sect5}, $H(\tau)$ changes
little (less than a factor 5) from $T_{BI}$ to $T_{EI}$, so that
$H_{BI}$ is its approximate scale throughout warm inflation.
Therefore, the thermalization condition Eq. (\ref{therm}) requires
$\alpha < 1$.

A stronger adiabatic condition also requires that the
$\varphi_0$ dependence in all Boltzmann factors varies slowly relative to the
thermalization rate. The relevant thermal excitations in this model are
the $\chi_{ki}$ and $\psi_{ki}$ fields when $m_{\chi_{ki},\psi_{ki}} <T$,
for which the stringest form
of the {\bf thermal-adiabatic condition} is
\begin{equation}
\frac{1}{T} \frac{d m_{\chi , \psi}}{d \tau} = 
\frac{g d\varphi_0 /d \tau}{T}  \ll
\frac{\Gamma_{\chi}(T)}{\Gamma_{\chi}(M)}.
\label{thermadiab}
\end{equation}
The estimates to follow are based on this criteria,
but it is much more restrictive than necessary
for two reasons.  {}First thermal excitation generally initiates
once $m_{\chi,\psi} \stackrel{<}{\sim} 10T$. 
{}For the DM-model this implies mass sites have quite a
long time to excite before their important role
arises, which is when $m_{\chi,\psi} \stackrel{<}{\sim} T$ 
at which point they drive the viscosity. 
Second, within the thermally excited
regime $m_{\chi,\psi} \stackrel{<}{\sim} T$, 
the Boltzmann factor is saturated at $\approx 1$, so
is insensitive to mass variations.

The next consistency check is the force condition.  In our case
this requires the $\lambda \varphi_0^3/6$ term to 
dominate the
effective equation of motion.  Due to vacuum energy cancellations
between bosons and fermions,  
the one-loop $T=0$ radiative corrections \cite{dj,mgleiser,sher}, 
$\sim \sum_i^{all} g^4 N (\varphi_0-M_i)^4$, 
$\sum_{i}^{all}g^4 N (\varphi_0-M_i)^4
\ln[(\varphi_0-M_i)^2/\mu^2_{\rm renorm}]$, are eliminated \cite{bgr2},
where the sum here is over all mass sites.. 
The T-dependent force terms from the effective potential are
1. $(g^2 N/8) T^2 \sum_i^{t.e.}(\varphi_0-M_i)$,
2. $(g^3 N/(4 \pi)) T \sum_i^{t.e.}(\varphi_0-M_i)|\varphi_0-M_i|$, and
3. $(g^4 N/(64 \pi)) \sum_i^{t.e.}(\varphi_0-M_i)^3 \log(T/\mu_{\rm renorm})$. 
Our approximation makes
a sharp division between thermally excited and unexcited sites. 
If the mass of a site 
$m_{\chi_{ik},\psi_{ik}}^2 \approx g^2(\varphi_0-M_i)^2 \leq T^2$,
that site is thermally excited and 
for $m_{\chi_{ik},\psi_{ik}}^2>T^2$
it is thermally unexcited or cold.  Here we assume
$\varphi_0$ is always surrounded by sufficiently many mass sites
on both sides so that there are no edge effects.  

There are two general features about all three types of force
terms.  {}First, the direction of the force flips for mass
sites on opposite sides of $\varphi_0$ and this implies
considerable cancellation in the summations.  Second, all
three forces are periodic in $\varphi_0$ with periodicity M.
As $\varphi_0$ traverses any M-interval, it experiences an
identical force profile in both directions.  Thus the average
force experienced by $\varphi_0$ from all three force terms
is zero after traversing any M-interval.

In our estimates, we use the most stringent consistency 
condition, which is to estimate the largest effect from these
force terms and demand that the $\lambda \varphi_0^3/6$ term
dominates it. We will find a large warm inflation solution
space with this stringent condition.
However estimates based on this force condition are overly
conservative, since  realistically $\varphi_0$ is not sharp, as we
are treating it, but rather is smeared over some interval
$\Delta \varphi_0$.   Smearing would imply the average force 
experienced by the $\varphi_0$-packet diminishes due to the
directionality dependence of the force terms.  In fact if
$\Delta \varphi_0 > M$, the force experienced by the wavepacket
from these three force terms
would be negligible.  Nevertheless, for simplicity
of the calculation, here this fact will not be treated and we will estimate
the the upper bound of the three force terms for a sharp $\varphi_0$.

Amongst the three force terms, note that 1 is the largest.
Ignoring the premultiplying factors (which also is largest for 1),
since for all thermally excited sites $g|\varphi_0-M_i| < T$, it
implies $\sum_i^{te}(\varphi_0-M_i) T^2 > 
\sum_i^{t.e.} g (\varphi_0-M_i) |\varphi_0-M_i| T >
\sum_i^{t.e.} g^2 (\varphi_0-M_i)^3$.
Thus provided $\lambda \varphi_0^3/6$ dominates force term 1, it
dominates them
all\footnote{When all bosons and
fermions are thermally excited, due to our choice of coupling
and of ratio between bose and fermi fields, term 3, in fact, cancels
amongst the bosons and fermions \cite{dj}.}.
Approximating
\begin{equation}
g \sum_i^{t.e.} (\varphi_0-M_i) \approx T
\end{equation}
we obtain the {\bf force condition}
\begin{equation}
\frac{\lambda \varphi_0^3}{6} > 
\frac{g^2N}{8} \sum_i^{t.e.} (\varphi_0-M_i) T^2
\approx \frac{gN}{8} T^3
\label{forcec}
\end{equation}

The final consistency check is the infra-red condition.
For the Minkowski space quantum field theory formalism
used here to be valid,
the Compton wavelength of all particle excitations should be
much smaller than the Hubble radius $1/H(\tau)$ during inflation.  {}For the
$\chi_{ik}$,$\psi_{ik}$ fields, this condition is easily satisfied since their
masses are generally $m_{\chi_{ik},\psi_{ik}} \approx T$ and
$T \gg H$.  This follows since for most of the time
the $\varphi_0(\tau)$-induced portion of their masses is large. 
However, even when $\varphi_0(\tau) \approx M_i$, the thermal mass
contribution $\sim gT$  generally is much larger than $H$.
On the other hand, the $\phi$-mass is more concerning, since the
self-coupling parameter $\lambda$ is usually tiny in warm inflation.
Thus the essential constraint that must be checked to enforce the
{\bf infra-red condition} is
\begin{equation}
m_{\phi}(\varphi_0,T) \gg H.
\label{irc}
\end{equation}

In total, there are five consistency conditions. The thermal-adiabatic,
$\varphi_0$-adiabatic and thermalization conditions all are adiabatic
conditions. The latter, equivalently stated, requires the cosmological 
expansion rate to be adiabatic relative to the particle production rate.
The force condition is not fundamentally required, but we have imposed it to
simplify the $\varphi_0$ equation of motion.

\subsection{Summary of the Parameters}
\label{subsect3C}

In Sects. \ref{sect4} and \ref{sect5} 
two cosmological parameters will be computed,
the number of e-folds
$N_e$ and amplitude of density perturbations $\delta \rho/\rho$,
in terms of the microscopic and thermodynamic parameters of the
model and one overall scale which will be the Planck scale
$M_p$.  Since the calculations are fairly detailed, for convenience here
all definitions of parameters and related terminology are summarized.

The following terminology is used in this paper:
\begin{center}
\begin{tabular}{p{1.5in} c p{4in}}
dissipative heat bath fields & - & fields $\chi_{ik}$, $\psi_{ik}$ that are
part of the heat reservoir and have significant effect on the dissipative
dynamics of the inflaton. \\
non-dissipative heat bath fields & - & fields 
$\chi^r_i$, $\psi^r_i$ that only are 
part of the heat reservoir and have no significant effect on the dissipative
dynamics of the inflaton. \\
mass site & - & ``location'' of dissipative heat bath fields in the DM model
specified by the mass parameter 
$M_i \equiv M i/g$, with $i$ denoting the number of the
site. \\
dissipative coefficient & - & the factor multiplying $d\varphi_0/dt$ in the
$\varphi_0$ equation of motion, 
$\sum_i^{t.e.}(\varphi_0-M_i)^2 \eta_{1i}^B(T)$ \\
dissipative function & - & $\eta_1^B(T) = \eta'g^4/(\pi T {\cal C})$
with $\eta'$ and ${\cal C}$ defined below Eq. (\ref{eta1B})
\end{tabular}
\end{center}

The microscopic parameters have all
been defined in detail in Sect. \ref{sect2}.  
Briefly, they are as follows:
\begin{center}
\begin{tabular}{p{1in} c p{4in}}
N & - &  number of $\chi_{ik}$ fields at every mass site $M_i$.
Correspondingly the number of $\psi_{ik}$ fields at every mass site is
$N/4$. \\
$N_r$ & - & number of nondissipative heat bath fields \\
$N_M$ & - & total number of mass sites crossed by the inflaton
$\varphi_0$ during warm inflation. Ignoring small corrections at the two
end points, this is equivalent to the total number of mass sites that
were thermally excited at some point during warm inflation. \\
$\lambda$ & - & $\phi$ self coupling parameter \\
$g$ & - & $\phi$ - $\chi$ coupling $\sim g^2/2$ , $\phi$ - $\psi$ coupling $\sim
g$ \\
$f$ & - & $\chi$ self coupling parameter  \\
$V_0$ & - & vacuum energy density shift parameter \\
$M$ & - & splitting scale between adjacent mass sites $g|M_{i+1} - M_i| =M$ \\
$m_{\chi_{ik}, \psi_{ik}}$ & - & mass of $\chi_{ik}$ or $\psi_{ik}$ fields
given in Eqs. (\ref{rmchi}), (\ref{rmpsi}) \\ 
$m_{\phi}$ & - &  $\phi$ mass given in Eq. (\ref{rmphi}) \\
\end{tabular}
\end{center}

The thermodynamics of the warm inflation is expressed through the
following quantities:
\begin{center}
\begin{tabular}{p{1.5in} c p{4in}}
$T_{BI}$ & - & temperature at the beginning of warm inflation \\
$T_{EI}$ & - & temperature at the end of warm inflation \\
$\kappa_M$ & - & $T_{BI}/M$ \\
$\beta$ & - & $T_{EI}/T_{BI}$ \\
\end{tabular}
\end{center}

The cosmology is expressed through the following quantities:
\begin{center}
\begin{tabular}{p{1.5in} c p{4in}}
$R$($\tau$) & - & scale factor, with $R(0) =1$ \\
$H_{BI}$ & - & Hubble parameter at the beginning of warm inflation,
$H_{BI} \equiv H(\tau_{BI})$. In the four cases examined in this paper,
the scale of $H(\tau)$ is of order $H_{BI}$ throughout the inflationary
period. \\
$\alpha$ & - & $H_{BI} \equiv \alpha \beta \kappa_M \Gamma_{chi}(M)$.
It will be seen that $\alpha$ is the adiabatic parameter with
adiabaticity increasing as $\alpha \rightarrow 0$. \\
$\rho_v(\tau)$  & - & vacuum energy density $\equiv V_0 +
\rho_{\phi}(\tau)$ where $\rho_{\phi}$ is the $\varphi_0$ dependent
portion. \\
$\rho_r(\tau)$ & - & radiation energy density \\
$N_e$ & - &  number of e-folds \\
$\delta \rho/\rho$ & - & amplitude of scalar density perturbation \\

\end{tabular}
\end{center}

Some additional notation used in this paper is as follows:
\begin{center}
\begin{tabular}{p{1.5in} c p{4in}}
BI & - & begin warm inflation \\
EI & - & end warm inflation \\
$\Gamma_{\chi}(M)$ & - & decay width for $\chi$-fields when they are
thermally excited, explicit expression in Eq. (\ref{chidecay}) \\
$\tau$ & - & rescaled time $\tau \equiv \Gamma_{\chi}(M) t$ \\
$n_{t.e.}$ & - & number of thermally excited ($t.e.$) mass sites \\
$\phi({\bf x}, \tau)$ & - & quantum inflaton field operator \\
$\varphi({\bf x}, \tau)$ & - & classical inflaton field
$\langle \phi({\bf x}, \tau) \rangle_T \equiv \varphi({\bf x},\tau)
=\varphi_0(\tau) + \delta \varphi({\bf x},\tau)$,
where $\varphi_0(\tau)$ is the homogeneous background field \\
$y$ & - & slope parameter for $\varphi_0$. Since the evolution is
overdamped, in all cases $d\varphi_0/d\tau \propto y \varphi_0$. \\
$k_F$ & - & freeze-out momentum for density perturbations, defined
Eq. (\ref{kfcond}) \\
$\Delta \varphi^2_H(\tau)$ & - & amplitude of scalar field fluctuations,
defined Eqs. (\ref{Dvp}) and (\ref{Dvplb}) \\
${\bf k}_p, {\bf k}_c$ & - & physical ($p$) and comoving ($c$)
wavenumber. By our convention they are the same at the end of warm
inflation $\tau_{EI}$, 
${\bf k}_p(\tau) \equiv {\bf k}_c/R(\tau -\tau_{EI})$. \\
$\delta \varphi_{{\bf k}_c}(\tau)$ & - & Fourier mode
of the inflaton field $\delta \varphi({\bf x},\tau)$; equivalently
this mode may be expressed as $\delta \varphi({\bf k}_p, \tau)$,
where the relation between ${\bf k}_p (\tau)$ and ${\bf k}_c$ is
specified. \\
\end{tabular}
\end{center}

{}Finally, for the following expressions, we simply state where they
are defined: $\eta' = 48 \ln(2T/m_{\chi}) \approx 48$
 - below Eq. (\ref{calc}); $\Gamma' \approx 1/192$ below Eq. (\ref{chidecay});
${\cal C}$ - Eq. (\ref{calc}), also
see first part of Sect. \ref{sect6}; $\Upsilon$ - Eq.(\ref{upsdef}).

\section{Solutions for e-folds $N_e$}
\label{sect4}

In this section four warm inflation solutions are presented.
The solutions we examine are for cases where
$T_{BI} \gg T_{EI}$ in Subsect. \ref{subsect4A} and
$T_{BI} \stackrel{>}{\sim} T_{EI}$ in Subsect. \ref{subsect4B}. 
We seek solutions in which $N_e$ e-folds of inflation occur with the
temperature of the Universe $T_{BI}$ and $T_{EI}$ respectively
at the beginning and end of warm inflation.
The temperatures $T_{BI}$ and $T_{EI}$ are
parameterized by $\beta$ and $\kappa_M$ in Eqs. (\ref{betadef})
and (\ref{kapdef}) respectively 
These conditions on the solutions imply two parameters of the model are
determined in terms of the others.
We choose these two
quantities to be $\lambda$ and $\varphi_{BI}/M$.  Under these
conditions the solutions for $\varphi_0(\tau)$ and $T(\tau)$ are
given as well as expressions for $\lambda$ and $\varphi_{BI}/M$. In
addition, for ease of reference, expressions are written for
$m_{\phi}^2/T^2$, $\rho_v(\tau)/\rho_r(\tau)$,
$d\varphi_0 /d \tau$ and $\Gamma_{\chi}(\tau)/\Gamma_{\chi}(M)$.
{}Finally the consistency conditions from Subsect. \ref{subsect3B}
are evaluated. The calculation
was designed so that the thermalization condition
Eq. (\ref{therm}) is satisfied simply by requiring $\alpha < 1$.
The adiabatic, force, and infra-red conditions
result in two essential inequalities that restrict the
self-consistent region of the parameter
space.  No overall scale is chosen in this section, since the
results leave this choice completely arbitrary.  

Although all the calculations are very simple, due to the generality of
the results, the final expressions may not appear transparent.  It
must
be appreciated that despite the compactness of the final expressions,
they contain the solutions under the very general situation in which
the parameters can be varied over a wide range and in which four scales
are adjustable, the initial temperature, the final temperature, the
Hubble expansion rate and the duration of warm inflation.

\subsection{$T_{BI} \gg T_{EI}$}
\label{subsect4A}

In this regime when the temperature from the beginning $T_{BI}$ 
to the end $T_{EI}$ of warm inflation 
changes significantly, the time dependence of the number of thermally
excited sites is treated in the $\varphi_0$-equation of motion. 
One general parametric constraint for this regime is
\begin{equation}
\beta \ll 1.
\end{equation}
At temperature $T(\tau)$, any $\chi_{ik}$  fields with mass
$m_{\chi_{ik}}^2 \approx g^2(\varphi_0-M_i)^2 < T^2$ is
thermally excited. Due to our choice of spacings between mass sites
$M_i$, this
implies approximately $T(\tau)/M$ sites 
adjacent to the field amplitude
$\varphi_0$ on either side are thermally excited.
In our approximation for all
sites with $m_{\chi_{ik}} > T$, the contribution to the
dissipative dynamics is ignored. Eq. (\ref{eom1})
therefore becomes
\begin{equation}
\frac{\eta' \Gamma'}{\pi^2}
N g^2 T^2 \frac{d \varphi_0}{d \tau} = 
- \frac{\lambda \varphi^3_0}{6}.
\label{eomA}
\end{equation}

Also, we allow for an additional $N_r$ bosonic
and $N_r/4$ fermionic non-dissipative heat bath fields
that contribute to the radiation  energy density but have no
affect on the dissipative dynamics of the inflaton.  
Thus the radiation energy density at any instant during warm
inflation is
\begin{equation}
\rho_r(\tau) = \frac{\pi^2}{16} 
[N_r + \frac{2NT(\tau)}{M}] T^4(\tau).
\label{rhora}
\end{equation}
We examine the solutions for the cases  $N_r=0$ and
$N_r \gg 2T_{BI}/T_{EI}$ in Subsects. \ref{subsubsect4A1}
and \ref{subsubsect4A2} respectively.
{}For both cases, the vacuum energy density shift parameter will be
$V_0 = \lambda \varphi^4_{BI}/24$, so that
the vacuum energy is approximately constant throughout warm inflation.
In this case the Hubble parameter,
$H \equiv \sqrt{8\pi G \rho} \approx \sqrt{8 \pi G \rho_v}$, 
changes by an O(1)-factor during warm inflation and
the number of e-folds is
\begin{equation}
N_e \approx H_{BI} t_{EI} = \alpha \beta \kappa_M \tau_{EI}.
\label{ne}
\end{equation}

\subsubsection{\bf $N_r=0$}
\label{subsubsect4A1}

In this case there are no non-dissipative heat bath fields.

\begin{center}
{\bf Solutions}
\end{center}

Eq. (\ref{dotr}) and Eq. (\ref{eomA}) imply
\begin{equation}
T(\tau) = \left( 
\frac{\lambda^2}{18 \eta' \Gamma' \alpha \beta \kappa_M N^2 g^2 }\right)^{1/7}
M^{1/7} \varphi_0^{6/7}(\tau).
\label{tempa1t}
\end{equation}
Substituting into Eq. (\ref{eomA}), we find the solution
\begin{equation}
\varphi_0(\tau) = \frac{\varphi_{BI}}{(y \tau +1)^{7/2}},
\label{phia1}
\end{equation}
where $y=[\pi^2 2^{2/7}/(3^{3/7} 7 (\eta'\Gamma')^{5/7})]
[(\alpha \beta \kappa_M)^2 \lambda^3/(N^3 g^{10})]^{1/7}
(\varphi_{BI}/M)^{2/7}$.
{}From
Eq. (\ref{ne}) for $N_e$ and  Eq. (\ref{betadef}) for $\beta$,
\begin{equation}
y=\frac{\alpha \beta^{2/3} \kappa_M}{N_e} (1-\beta^{1/3}).
\end{equation}
By equating this expression for $y$ with the one in terms of the
parameters in the model, and applying Eq. (\ref{kapdef}) to
eq. (\ref{tempa1t}), two parameters of the model are determined, of which
we choose
\begin{equation}
\lambda= \left(\frac{1029 \eta^{'2} \Gamma^{'2}}{2 \pi^6}\right)
\frac{\alpha^2 \beta \kappa_M (1-\beta^{1/3})^3 N g^4}
{N_e^3}
\label{lama1}
\end{equation}
and
\begin{equation}
\frac{\varphi_{BI}}{M}= (18\eta'\Gamma')^{1/6}
\left(\frac{\alpha \beta N^2g^2}{\lambda^2}\right)^{1/6} \kappa_M^{4/3}
=\left(\frac{\pi^2}{7}\sqrt{\frac{2}{\eta'\Gamma'}}
\right)
\frac{\kappa_M N_e}
{\alpha^{1/2} \beta^{1/6} g (1-\beta^{1/3})}.
\label{phima1}
\end{equation}

Using Eqs. (\ref{lama1}) and (\ref{phima1}), Eq. (\ref{tempa1t}) becomes
\begin{equation}
T(\tau) = \left( \frac{49 \eta' \Gamma'}{2 \pi^4} \right)^{3/7}
\left(\frac{\alpha^{3/7} \beta^{1/7} \kappa_M^{1/7}(1-\beta^{1/3})^{6/7}
g^{6/7}}{N_e^{6/7}} \right) M^{1/7} \varphi_0^{6/7}(\tau).
\label{tempa1}
\end{equation}
Eqs. (\ref{phia1}) and (\ref{tempa1}) are the general solutions.

Based on this solution, next some useful expressions are given.
The number of sites that are thermally excited at a given instance
is $n_{te} \equiv 2T(\tau)/M = 2\kappa_M/(y \tau+1)^3$.
The number of mass sites $\varphi_0$ crosses during the warm inflation
period is
\begin{equation}
N_M \equiv \frac{g|\varphi_{BI} -\varphi_{EI}|}{M}
= \frac{\pi^2}{7}\sqrt{\frac{2}{\eta'\Gamma'}}
\frac{\kappa_M N_e (1-\beta^{7/6})}{\alpha^{1/2} \beta^{1/6} (1-\beta^{1/3})} 
\end{equation}
with
\begin{equation}
i_{\rm min} \equiv \frac{g \varphi_{EI}}{M} =
\frac{\pi^2}{7}\sqrt{\frac{2}{\eta'\Gamma'}}
\frac{\beta \kappa_M N_e}{\alpha^{1/2} (1-\beta^{1/3})}.
\end{equation}
The mass of $\phi$ is
\begin{equation}
{\bar m}_{\phi}^2 \equiv \frac{m_{\phi}^2}{T^2} 
\approx \frac{\lambda \varphi_0^2}{2T^2}=
\left(\frac{21 \eta'\Gamma'}{2 \pi^2} \right)
\frac{\alpha \beta^{2/3}(1-\beta^{1/3}) \kappa_M N g^2}{N_e (y \tau+1)}.
\end{equation}
The ratio of vacuum to radiation energy density is
\begin{equation}
\frac{\rho_{\phi}(\tau)}{\rho_r(\tau)} = 
\frac{2\beta^{1/3}N_e}{7(1-\beta^{1/3})}(y \tau+1)
\label{rhorvA1}
\end{equation}
\begin{equation}
\frac{V_0}{\rho_r(\tau)} = 
\frac{2\beta^{1/3}N_e}{7(1-\beta^{1/3})}(y \tau+1)^{15}.
\end{equation}
{}Finally, for examining the adiabatic condition
\begin{equation}
\frac{d \varphi_0}{d \tau} = 
\frac{7 \alpha \beta^{2/3} \kappa_M (1-\beta^{1/3})}{2N_e}
\frac{\varphi_0(\tau)}{(y \tau+1)} =
\left(\frac{\pi^2}{\sqrt{2\eta'\Gamma'}} \right)
\frac{\alpha^{1/2} \beta^{1/2} \kappa_M}{g}
\frac{T(\tau)}{(y \tau+1)^{3/2}}
\label{dphia1}
\end{equation}
and
\begin{equation}
\frac{\Gamma_{\chi}(T)}{\Gamma_{\chi}(M)} =
\frac{\kappa_M}{(y \tau+1)^3}.
\label{therma1}
\end{equation}

\begin{center}
{\bf Consistency Conditions}
\end{center}

The above solutions still are subject to consistency checks. 
{}For the $\varphi_0$-adiabatic condition Eq. (\ref{padiab}), 
since the thermalization rate
Eq. (\ref{therma1}) decreases faster 
than $(d\varphi_0(\tau)/d\tau)/\varphi_0$ from Eqs. 
(\ref{phia1}) and (\ref{dphia1}), the most stringent test is at $\tau_{EI}$;
if the $\varphi_0$-adiabatic condition holds at $\tau_{EI}$, then it is
satisfied better at earlier times.  We find the
$\varphi_0$-adiabatic condition Eq. (\ref{padiab})
requires $7\alpha/(2N_e) < 1$.  Since $\alpha<1$, for this to hold,
it is sufficient that
\begin{equation}
N_e > 7/2.
\end{equation}

The thermal-adiabatic condition
Eq. (\ref{thermadiab}) implies from Eqs. (\ref{tempa1}),
(\ref{dphia1}) and (\ref{therma1}) that the most
stringent time is $\tau_{EI}$ with the constraint
\begin{equation}
\frac{\pi^2}{\sqrt{2\eta'\Gamma'}} \alpha^{1/2} < 1.
\label{thermaA1}
\end{equation}
By substituting for $\eta'$ and $\Gamma'$, it gives
$\alpha < 1/199 $ (at $\tau=0$ the condition implies an additional factor
of $\beta^{1/2}$ in Eq. (\ref{thermaA1}) ).

The force
condition Eq. (\ref{forcec}) implies
\begin{equation}
4\sqrt{2\eta'\Gamma'} \alpha^{1/2} \beta \kappa_M > 1.
\label{forcecA1}
\end{equation}

{}Finally the infra-red condition Eq. (\ref{irc}) implies
\begin{equation}
\frac{21 \eta'}{2 \Gamma'} \frac{\beta (1-\beta^{1/3}) \kappa_M N g^2}
{\alpha N_e (g^4+f^2/8)^2} > 1.
\end{equation}
{}For the force and infra-red conditions, they also are
evaluated at the most stringent instant during warm
inflation, which turns out to be again $\tau=\tau_{EI}$.

\subsubsection{\bf $N_r \gg 2NT/T_{EI}$}
\label{subsubsect4A2}

In this case the $N_r$ term dominates the radiation energy density in Eq.
(\ref{rhora}).  

\begin{center}
{\bf Solutions}
\end{center}

The procedure for solving this case is similar to the above
case. The results are
\begin{equation}
\varphi_0(\tau) = \varphi_{BI} \exp(-y \tau)
\end{equation}
and
\begin{equation}
T(\tau) = \frac{1}{3^{1/3} (\eta' \Gamma')^{1/6}}
\left(\frac{\lambda^2}{\alpha \beta \kappa_M N_rN g^2} \right)^{1/6} 
\varphi_0(\tau) = \kappa_M M \exp(-y\tau),
\label{tempA2}
\end{equation}
where in terms of the parameters of the model
$y=[\pi^2/(3^{1/3} 2 (\eta' \Gamma')^{2/3})][\lambda N_r \alpha \beta
\kappa_M/(N^2 g^4)]^{1/3}$ and in terms of the expansion e-folds
from Eq. (\ref{ne})
\begin{equation}
y= \frac{\alpha \beta \kappa_M \ln(1/\beta)}{N_e}.
\end{equation}
{}From the specified conditions on temperature expressed through
$\alpha$, $\beta$, and $\kappa_M$, two parameters in the model are
determined 
\begin{equation}
\lambda = \left(\frac{24 \eta^{'2}\Gamma^{'2}}{\pi^6}\right)
\frac{\alpha^2 \beta^2 \kappa_M^2 \ln^3(1/\beta) N^2 g^4}{N_e^3 N_r}
\end{equation}
and
\begin{equation}
\frac{\varphi_{BI}}{M} = 3^{1/3}(\eta' \Gamma')^{1/6}
\left(\frac{\alpha \beta \kappa_M N_r N g^2}{\lambda^2}\right)^{1/6}
\kappa_M =\frac{\pi^2}{2(\eta'\Gamma')^{1/2}} 
\frac{\kappa_M^{1/2} N_eN_r^{1/2}}{\alpha^{1/2} \beta^{1/2}
\ln(1/\beta) N^{1/2} g}.
\end{equation}
Based on these solutions, some useful expressions are as follows.
The number of mass sites that are thermally excited at a given instance
are $n_{t.e.} = 2 \kappa_M \exp(-y \tau)$.  The number of mass sites that
$\varphi_0$ crosses during the warm inflation period is
\begin{equation}
N_M=\frac{\pi^2}{2(\eta'\Gamma')^{1/2}} 
\frac{(1-\beta)\kappa_M^{1/2} N_e N_r^{1/2}}{\alpha^{1/2}\beta^{1/2}
\ln(1/\beta) N^{1/2}},
\end{equation}
with
\begin{equation}
i_{\rm min} =
\frac{\pi^2}{2(\eta'\Gamma')^{1/2}} 
\frac{\beta^{1/2}\kappa_M^{1/2} N_e N_r^{1/2}}{\alpha^{1/2}
\ln(1/\beta) N^{1/2}}.
\end{equation}
The mass of the $\phi$ particles is
\begin{equation}
{\bar m}_{\phi}^2 \approx \frac{\lambda \varphi_0^2}{2 T^2}
=\frac{3\eta'\Gamma'}{\pi^2} 
\frac{\alpha \beta \kappa_M \ln(1/\beta) N g^2}{N_e}.
\end{equation}
The ratio of vacuum to radiation energy density is
\begin{equation}
\frac{\rho_{\phi}(\tau)}{\rho_r(\tau)}=
\frac{1}{(\eta'\Gamma')^{1/2}} \frac{N_e}{\ln(1/\beta)}
\label{rhorvA2}
\end{equation}
\begin{equation}
\frac{V_0}{\rho_r(\tau)}=
\frac{1}{(\eta'\Gamma')^{1/2}} \frac{N_e}{\ln(1/\beta)}
\exp(4y\tau).
\end{equation}
{}For examining the adiabatic conditions
\begin{equation}
\frac{d \varphi_0}{d \tau} = 
-\frac{\alpha \beta \ln(1/\beta) \kappa_M}{N_e} \varphi_0(\tau)
=\frac{\pi^2}{2(\eta'\Gamma')^{1/2}} 
\frac{\alpha^{1/2} \beta^{1/2} \kappa_M^{1/2} N_r^{1/2}}
{N^{1/2} g} T(\tau)
\end{equation}
and
\begin{equation}
\frac{\Gamma_{\chi}(T)}{\Gamma_{\chi}(M)} =
\kappa_M \exp(-y \tau).
\end{equation}

\begin{center}
{\bf Consistency Conditions}
\end{center}

The parametric constraints from the consistency conditions are
the $\varphi_0$-adiabatic condition Eq. (\ref{padiab}) with sufficiency
condition
\begin{equation}
N_e > 1,
\end{equation}
the thermal-adiabatic condition Eq. (\ref{thermadiab})
\begin{equation}
\frac{\pi^2}{2(\eta'\Gamma')^{1/2}} 
\left(\frac{\alpha N_r}{\beta \kappa_M N}\right)^{1/2} < 1,
\label{thermaA2}
\end{equation}
the force condition Eq. (\ref{forcec})
\begin{equation}
4(\eta'\Gamma')^{1/2} \left(\frac{\alpha \beta \kappa_M
N_r}{N}\right)^{1/2} > 1,
\label{forcecA2}
\end{equation}
and the infrared condition Eq. (\ref{irc})
\begin{equation}
\frac{3\eta'}{\Gamma'} \frac{\beta \ln(1/\beta) \kappa_M N g^2}
{\alpha N_e (g^4+f^2/8)^2} > 1.
\end{equation}
It should also be recalled that the basic requirement for
this regime, $N_r$-dominated, is
\begin{equation}
\frac{N_r}{N} \gg 2 \kappa_M. 
\end{equation}

\subsection{$T_{BI} \stackrel{>}{\sim} T_{EI}$}
\label{subsect4B}

{}For this regime, we adopt the criteria 
\begin{equation}
\beta \stackrel{>}{\sim} 0.5.
\label{beta5}
\end{equation}
In this regime
the number of thermally excited sites in the $\varphi_0$-equation of motion
is approximated as constant, so that
in Eq. (\ref{eom1}) $\sum_i^{t.e.}(\varphi_0-M_i)^2 \eta_{1i}^B(T)
\approx \kappa_M^3 M^2 N \eta'/(g^2T)$.   All other approximations are the same
as in the previous section. Thus the $\varphi_0$-equation of motion is
\begin{equation}
\frac{\eta'\Gamma'\kappa_M^3 g^2 N M^3 }{\pi^2 T}
\frac{d \varphi_0}{d \tau} = -\frac{\lambda \varphi_0^3}{6}.
\label{eomB}
\end{equation}
Both cases studied below are for arbitrary number of dissipative and
non-dissipative heat bath fields so that the radiation energy density is
\begin{equation}
\rho_r(\tau) = \frac{\pi^2}{16} 
(N_r + 2 N \kappa_M) T^4(\tau).
\end{equation}
The two cases examined in order are $V_0 >0$ and $V_0=0$.
The latter case is the DM-model examined in \cite{bgr2}.  However here
the calculation is extended to permit an arbitrary mass splitting scale
M (in \cite{bgr2} the mass splitting scale was
restricted to $M \sim T$)
and to treat the time dependence of the temperature in the
$\varphi_0$-equation of motion.

\subsubsection{$V_0> 0$}
\label{subsubsect4B1}

Similar to the previous subsection, we choose
$V_0 \approx \lambda \varphi^4_{BI}/24$. The number of e-folds
follows from the relation Eq. (\ref{ne}).

\begin{center}
{\bf Solutions}
\end{center}

The procedure for solving this case is similar to the above
cases. The results are
\begin{equation}
\varphi_0(\tau) = \frac{\varphi_{BI}}{(y\tau+1)^{1/4}}
\end{equation}
and
\begin{equation}
T(\tau) = \frac{1}{(9 \eta' \Gamma')^{1/3}}
\left(\frac{\lambda^2}{\alpha \beta \kappa_M^4 (N_r+2\kappa_M N) N g^2} 
\right)^{1/3} 
\frac{\varphi_0(\tau)^2}{M} =
\frac{\kappa_M M}{(y \tau+1)^{1/2}},
\label{tempB1}
\end{equation}
where in terms of the parameters of the model
$y=[2\pi^2/(3^{5/3}(\eta' \Gamma')^{4/3}]
[\lambda^5/(\alpha \beta \kappa_M^{13} (N_r+2\kappa_M N) N^4 g^8]^{1/3}
(\varphi_{BI}/M)^4$
and in terms of the expansion e-folds
from Eq. (\ref{ne})
\begin{equation}
y= \frac{\alpha \kappa_M (1-\beta^2)}{\beta N_e}.
\end{equation}
{}From the specified conditions of temperature expressed through
$\alpha$, $\beta$, and $\kappa_M$ two parameters in the model are
determined 
\begin{equation}
\lambda = \left(\frac{3 \eta^{'2}\Gamma^{'2}}{8\pi^6}\right)
\frac{\alpha^2 (1-\beta^2)^3 \kappa_M^2 N^2 g^4}
{\beta^4 N_e^3 (N_r+2\kappa_M N) }
\end{equation}
and
\begin{equation}
\frac{\varphi_{BI}}{M} = (9 \eta' \Gamma')^{1/6}
\left(\frac{\alpha^2 \beta^2 \kappa_M^7 (N_r+2\kappa_M N)  N g^2}
{\lambda^2}\right)^{1/6} =
\frac{2 \pi^2}{(\eta'\Gamma')^{1/2}} 
\frac{\beta^{3/2} \kappa_M^{1/2} N_e (N_r+2\kappa_M N)^{1/2}}
{\alpha^{1/2} (1-\beta^2) N^{1/2} g}.
\end{equation}
Based on these solutions, some useful expressions are as follows.
The number of mass sites that are thermally excited at a given instance
are $n_{t.e.} = 2 \kappa_M/(y \tau+1)^{1/2}$.  The umber of mass sites that
$\varphi_0$ crosses during the warm inflation period is
\begin{equation}
N_M=\frac{2 \pi^2}{(\eta'\Gamma')^{1/2}} 
\frac{\beta^{3/2}(1-\beta^{1/2})\kappa_M^{1/2} N_e 
(N_r+2\kappa_M N)^{1/2}}{\alpha^{1/2}(1-\beta^2) N^{1/2}},
\end{equation}
with
\begin{equation}
i_{\rm min} =
\frac{2 \pi^2}{(\eta'\Gamma')^{1/2}} 
\frac{\beta^2\kappa_M^{1/2} N_e 
(N_r+2\kappa_M N)^{1/2}}{\alpha^{1/2}(1-\beta^2) N^{1/2}}.
\end{equation}
The mass of the $\phi$ particles is
\begin{equation}
{\bar m}_{\phi}^2 \approx \frac{\lambda \varphi_0^2}{2 T^2}
=\frac{3\eta'\Gamma'}{4 \pi^2} 
\frac{\alpha \kappa_M (1-\beta^2) N g^2}{\beta N_e}(y\tau+1)^{1/2}.
\end{equation}
The ratio of vacuum to radiation energy density is
\begin{equation}
\frac{\rho_{\phi}(\tau)}{\rho_r(\tau)}=
\frac{4}{3^{4/3} (\eta'\Gamma')^{2/3}} \frac{\beta^2 N_e}{(1-\beta^2)}
(y\tau+1)
\label{rhorvB1}
\end{equation}
\begin{equation}
\frac{V_0}{\rho_r(\tau)}=
\frac{4}{3^{4/3} (\eta'\Gamma')^{2/3}} \frac{\beta^2 N_e}{(1-\beta^2)}
(y\tau+1)^2.
\end{equation}
{}For examining the adiabatic conditions
\begin{equation}
\frac{d \varphi_0}{d \tau} = 
-\frac{\alpha (1-\beta^2) \kappa_M}{4 \beta N_e} 
\frac{\varphi_0(\tau)}{(y \tau+1)}
=\frac{\pi^2}{2(\eta'\Gamma')^{1/2}} 
\frac{\alpha^{1/2} \beta^{1/2} \kappa_M^{1/2} (N_r+2\kappa_M N)^{1/2}}
{N^{1/2}g(y\tau+1)^{3/4}} T(\tau)
\end{equation}
and
\begin{equation}
\frac{\Gamma_{\chi}(T)}{\Gamma_{\chi}(M)} =
\frac{\kappa_M}{(y\tau+1)^{1/2}}.
\end{equation}

\begin{center}
{\bf Consistency Conditions}
\end{center}

The parametric constraints from the consistency conditions are
the $\varphi_0$-adiabatic condition 
Eq. (\ref{padiab}) with sufficiency condition
\begin{equation}
N_e > \frac{3}{8},
\end{equation}
the thermal-adiabatic condition Eq. (\ref{thermadiab})
\begin{equation}
\frac{\pi^2}{2(\eta'\Gamma')^{1/2}} 
\left(\frac{\alpha \beta (N_r+2\kappa_MN)}{\kappa_M N}\right)^{1/2}
<1,
\label{thermaB1}
\end{equation}
the force condition Eq. (\ref{forcec})
\begin{equation}
4(\eta'\Gamma')^{1/2} \left(\frac{\alpha \beta \kappa_M
(N_r + 2\kappa_M N)}{N}\right)^{1/2} > 1,
\label{forcecB1}
\end{equation}
and the infrared condition Eq. (\ref{irc})
\begin{equation}
\frac{3\eta'}{4\Gamma'} \frac{(1-\beta^2) \kappa_M N g^2}
{\alpha \beta^2 N_e (g^4+f^2/8)^2} > 1.
\end{equation}

\subsubsection{$V_0 = 0$}
\label{subsubsect4B2}

In this case, there is no shift parameter $V_0$ nor extra function
$V_1(\phi)$ in the Lagrangian Eq. (\ref{Nfields}). This is the model
examined in \cite{bgr2}, except here the calculation is extended to treat
the time dependence of the temperature in the dissipative function. 
Also \cite{bgr2} only examined the regime where the mass level splitting
$M\sim T$, whereas here the relation between $M$ and $T$ can be varied
through the parameter $\kappa_M$.

There are two differences in this calculation's procedures
compared to the previous three cases.  
Both differences arise because here the Hubble
parameter is $\varphi_0(\tau)$ dependent 
\begin{equation}
H(\tau)=\alpha \beta \kappa_M 
\Gamma_{\chi}(M) \frac{\varphi_0^2(\tau)}{\varphi_{BI}^2}.
\label{hubve0}
\end{equation}
{}First this dependence must be treated in the energy conservation
equation (\ref{dotr}). Second the scale factor no longer grows exactly
exponentially. However this calculation is also
based on the assumption that  the temperature during warm
inflation does not change significantly, 
$\beta \stackrel{>}{\sim} 0.5$, and 
as will be seen this also implies $\varphi_0(\tau)$, thus $\rho_v(\tau)$,
does not change significantly.  In this case, the scale factor grows
quasi-exponentially with e-folds
\begin{equation}
N_e \approx \int_0^{t_{EI}} H(t) dt = 
\alpha \beta \kappa_M \int_0^{\tau_{EI}} \frac{\varphi_0^2(\tau)}
{\varphi_{BI}^2}
d\tau.
\label{nev00}
\end{equation}

\begin{center}
{\bf Solutions}
\end{center}

The procedure for solving this case is similar to the above
case. The results are
\begin{equation}
\varphi_0(\tau) = \frac{\varphi_{BI}}{(y\tau+1)^{3/10}}
\end{equation}
and
\begin{equation}
T(\tau) = \frac{1}{(9 \eta' \Gamma')^{1/3}}
\left(\frac{\lambda^2}{\alpha \beta \kappa_M^4 (N_r+2\kappa_M N) N g^2} 
\right)^{1/3} \left(\frac{\varphi_{BI}}{M}\right)^{2/3}
\frac{\varphi_0^{4/3}(\tau)}{M^{1/3}} =
\frac{\kappa_M M}{(y \tau+1)^{2/5}},
\label{tempB2}
\end{equation}
where in terms of the parameters of the model
$y=[5 \pi^2/(( 9\eta' \Gamma')^{4/3}]
[\lambda^5/(\alpha \beta \kappa_M^{13} (N_r+2\kappa_M N) N^4 g^8)]^{1/3}
(\varphi_{BI}/M)^4$
and in terms of the expansion e-folds
from Eq. (\ref{nev00})
\begin{equation}
y= \frac{5 \alpha \kappa_M (1-\beta)}{2 N_e}.
\end{equation}
{}From the specified conditions of temperature expressed through
$\alpha$, $\beta$, and $\kappa_M$ two parameters in the model are
determined 
\begin{equation}
\lambda = \left(\frac{81 \eta^{'2}\Gamma^{'2}}{8\pi^6}\right)
\frac{\alpha^2 (1-\beta)^3 \kappa_M^2 N^2 g^4}
{\beta N_e^3 (N_r+2\kappa_M N) }
\label{lamB2}
\end{equation}
and
\begin{equation}
\frac{\varphi_{BI}}{M} = 3 (\eta' \Gamma')^{1/2}
\left(\frac{\alpha \beta \kappa_M^7 (N_r+2\kappa_M N)  N g^2}
{\lambda^2}\right)^{1/6} =
\frac{2 \pi^2}{3(\eta'\Gamma')^{1/2}} 
\frac{\beta^{1/2} \kappa_M^{1/2} N_e (N_r+2\kappa_M N)^{1/2}}
{\alpha^{1/2} (1-\beta) N^{1/2} g}.
\end{equation}
Based on these solutions, some useful expressions are as follows.
The number of mass sites that are thermally excited at a given instance
are $n_{t.e.} = 2 \kappa_M/(y \tau+1)^{2/5}$.  The number of mass sites that
$\varphi_0$ crosses during the warm inflation period is
\begin{equation}
N_M=\frac{2 \pi^2}{3(\eta'\Gamma')^{1/2}} 
\frac{\beta^{1/2}(1-\beta^{3/4})\kappa_M^{1/2} N_e 
(N_r+2\kappa_M N)^{1/2}}{\alpha^{1/2}(1-\beta) N^{1/2}},
\end{equation}
with
\begin{equation}
i_{\rm min} =
\frac{2 \pi^2}{3(\eta'\Gamma')^{1/2}} 
\frac{\beta^{5/4}\kappa_M^{1/2} N_e 
(N_r+2\kappa_M N)^{1/2}}{\alpha^{1/2}(1-\beta) N^{1/2}},
\end{equation}
The mass of the $\phi$ particles is
\begin{equation}
{\bar m}_{\phi}^2 \approx \frac{\lambda \varphi_0^2}{2 T^2}
=\frac{9 \eta'\Gamma'}{4 \pi^2} 
\frac{\alpha (1-\beta) \kappa_M N g^2}{N_e}(y\tau+1)^{1/5}.
\end{equation}
The ratio of vacuum to radiation energy density is
\begin{equation}
\frac{\rho_{\phi}(\tau)}{\rho_r(\tau)}=
\frac{4}{3} \frac{\beta N_e}{(1-\beta)^4}
(y\tau+1)^{2/5}.
\label{rhorvB2}
\end{equation}
{}For examining the adiabatic conditions
\begin{equation}
\frac{d \varphi_0}{d \tau} = 
-\frac{3 \alpha (1-\beta) \kappa_M}{4 N_e} 
\frac{\varphi_0(\tau)}{(y \tau+1)}
=\frac{\pi^2}{2(\eta'\Gamma')^{1/2}} 
\frac{\alpha^{1/2} \beta^{1/2} \kappa_M^{1/2} (N_r+2\kappa_M N)^{1/2}}
{N^{1/2} g(y\tau+1)^{1/2}} T(\tau)
\end{equation}
and
\begin{equation}
\frac{\Gamma_{\chi}(T)}{\Gamma_{\chi}(M)} =
\frac{\kappa_M}{(y\tau+1)^{2/5}}.
\end{equation}

\begin{center}
{\bf Consistency Conditions}
\end{center}

The parametric constraints from the consistency conditions are
the $\varphi_0$ adiabatic condition Eq. (\ref{padiab}) 
with sufficiency condition
\begin{equation}
N_e > \frac{3}{8} ,
\end{equation}
the thermal-adiabatic condition Eq. (\ref{thermadiab})
\begin{equation}
\frac{\pi^2}{2(\eta'\Gamma')^{1/2}} 
\left(\frac{\alpha \beta (N_r+2\kappa_MN)}{\kappa_M N}\right)^{1/2}
<1,
\label{thermaB2}
\end{equation}
the force condition Eq. (\ref{forcec})
\begin{equation}
4(\eta'\Gamma')^{1/2} \left(\frac{\alpha \beta \kappa_M
(N_r + 2\kappa_M N)}{N}\right)^{1/2} > 1,
\label{forcecB2}
\end{equation}
and the infrared condition Eq. (\ref{irc})
\begin{equation}
\frac{9\eta'}{4\Gamma'} \frac{(1-\beta) \kappa_M N g^2}
{\alpha N_e (g^4+f^2/8)^2} > 1.
\end{equation}

\subsection{Discussion}
\label{subsect4C}

In this subsection, some general comments are given about the four cases
examined in the previous two subsections.  The first noteworthy
observation is that for arbitrary e-folds, $N_e$, the consistency
conditions impose very mild restrictions on the parameter space in all
four cases.  Although, the precise restrictions 
vary amongst the four
cases, a general set of restrictions that is valid for all four cases
can be given.  {}First recall that by construction of the solution,
the thermalization condition Eq(\ref{therm}) 
always requires simply that $\alpha < 1$.
The most stringent restrictions arise from the force Eq. (\ref{forcec}) and
thermal-adiabatic Eq. (\ref{thermadiab}) conditions, which together require
\begin{equation}
\frac{1}{\beta \kappa_M^2} < \alpha \stackrel{<}{\sim}
\frac{\beta \kappa_M N}{100(N_r+2\kappa_M N)}.
\label{genalp}
\end{equation}
The other two consistency conditions impose very mild restrictions. The
$\varphi_0$-adiabatic condition Eq. (\ref{padiab}) is 
always accommodated provided
$N_e>1$ and the infra-red condition Eq. (\ref{irc})
is accommodated provided
$\beta \kappa_M Ng^2/(\alpha N_e(g^4+f^2/8)^2 > 4 \times 10^{-5}$.
With all the conditions combined, they are fairly 
unrestrictive to the parameter space.  As such, it
leaves considerable freedom for treating density perturbations.

Another interesting point is to compare the 
solution in Subsect. \ref{subsubsect4B1}
for $T_{EI} \stackrel{<}{\sim} T_{BI}$ 
with those in Subsect. \ref{subsect4A} for $T_{EI} \ll
T_{BI}$ and see how well they match at some intermediate $T_{EI} <
T_{BI}$.  There clearly should be some overlapping region since the
model is exactly the same and only the treatment of temperature
dependence is different.  By comparing the expressions for $\lambda$ and
$\varphi_{BI}/M$, the functional dependence on the parameters
is seen to be exactly the same except for with respect to
$\beta$.  In regards to $\beta$, both $\lambda$ and $\varphi_{BI}/M$ 
for the cases in Subsects. \ref{subsubsect4A1}, $N_r=0$, and 
\ref{subsubsect4A2}, $N_r \gg 2T/M$,
equate to the expressions in \ref{subsubsect4B1} at $\beta=0.42$ and $0.53$
respectively.  This is close to the approximate cut-off criteria we gave
in 3B of $\beta \sim 0.5$ Eq. (\ref{beta5}).

The final point is that 
cross comparison amongst the four cases indicates several similarities amongst
the solutions.  In the remainder of this subsection,
the origin of these similarities
are examined. Alongside this, a qualitative understanding of the
solutions are developed.  

The two basic equations of warm inflation, the $\varphi_0$-equation of
motion Eq. (\ref{eom1}) and the energy conservation equation
Eq. (\ref{dotr}) have five properties,  which are listed below.
{}From the properties, all the similarities amongst the solutions then
are explained.  

\noindent
{\bf property 1}: The first time derivative of $\varphi_0$, 
$d\varphi_0(\tau)/d\tau$, is related to some product of 
$\varphi_0(\tau)^a T(\tau)^b$ in both basic equations of warm inflation,
Eqs. (\ref{eom1}) and (\ref{dotr}). (In the $\varphi_0$-equation of
motion, Eq. (\ref{eom1}), recall that we approximate the dissipative
coefficient as $\sum_i^{t.e.} (\varphi_0-M_i)^2 \eta(T)
\approx T^3  \eta(T)/(g^2 M)$, where for
the $T^3(\tau)$ term, in Subsect. A the time dependence is treated 
and in Subsect. B it is treated as a constant $T^3(\tau)=T^3(0)$.
In particular the dissipative coefficient in both subsections
depends on some power of the
temperature $T(\tau)$.)

\noindent
{\bf property 2}: In the $\varphi_0$-equation of motion, the direction of
$d\varphi_0(\tau)/d\tau$ is always opposite to the sign of $\varphi_0(\tau)$.

\noindent
{\bf property 3}: Since the $\varphi_0$-equation of motion is first order,
the solution requires one initial condition which then sets the overall scale
for both $\varphi_0(\tau)$ and $T(\tau)$.  In our approach, this scale is
set by the condition $T(0)\equiv T_{BI}=\kappa_M M$. As such $\varphi_{BI}$
is then a derived quantity. 

\noindent
{\bf property 4}: The force term in the $\varphi_0$ equation of motion 
in our approximation is
always $\lambda \varphi_0^3/6$.

\noindent
{\bf property 5}: The number of e-folds is linearly related to the
dimensionless time parameter at the end of warm inflation $\tau_{EI}$
in all four cases $N_e \propto \alpha \tau_{EI}$.

Properties one and two imply that the solutions have the general behavior
\begin{equation}
\varphi_0(\tau)
\equiv \varphi_{BI} D(\tau)
\label{varphigs}
\end{equation}
where
\begin{equation}
D(\tau) =
\frac{1}{(y\tau+1)^{\gamma_{\phi}}} 
( \ \ {\rm or} \ \ \exp(-y\tau) )
\end{equation}
and
\begin{equation}
T(\tau) = f(\lambda,g,f,N,\kappa_M,\alpha,\beta)
\varphi^{\gamma_T}_0(\tau) M^{1-\gamma_T}
\label{T1}
\end{equation}
with $\gamma_{\phi},\gamma_T >0$ and 
$f(\lambda,g,f,N,\kappa,\alpha,\beta)$ a function of the
the model and thermodynamic parameters.  
Combining this deduction with property 3, we also can conclude that in
general\footnote{Equating
Eqs. (\ref{T1}) and (\ref{T2}) lead to one of
the two parametric constraints that in the previous two subsections
determined $\lambda$ and 
$\varphi_{BI}/M$.}
\begin{equation}
T(\tau) = \kappa_M D^{\gamma_T}(\tau)
\label{T2}
\end{equation}
The noteworthy point for the present discussion is the solutions for
$\varphi_0(\tau)$ and $T(\tau)$ are always a product of a mass-dimension
one function which depends on the model and thermodynamic parameters
and a time dependent decay function. The latter we represent through
$D(\tau)$ taken to some positive power with $D(0)=1$ and 
$D(\tau>0) < 1$.  {}For this discussion it is useful to think about the
solutions this way, since the general features are contained in the
mass-dimension one function.  As such, the discussion to follow is not
detailed about the decay function.  

The general behavior of the temperature permits two deductions. {}First
from Eq.(\ref{T2}) and the definition of $\beta$ it follows that
in general
\begin{equation}
y \sim \frac{\alpha h(\beta) \kappa_M}{N_e}
\end{equation}
where $h(\beta)$ depends on the specific time dependence of $T(\tau)$
\footnote{Equating this expression for $y$ with the one obtained with
respect to the parameters of the model yields the second of two
conditions that determine $\lambda$ and $\varphi_{BI}/M$ in
the previous two subsections.}. 
Second, our approximation for the dissipative coefficient in all the cases
studied has the general form
\begin{equation}
\sum_i^{t.e.}(\varphi_0-M_i)^2 \frac{\eta'}{\pi T}
\approx \frac{T^3 \eta'}{\pi M T} \approx
\frac{\kappa_M^2 M \eta'}{\pi} D(\tau)^{\gamma_{\eta}},
\label{dissgen}
\end{equation}
where $\gamma_{\eta}$ depends on the specific treatment of the
temperature's time dependence. In Subsect. A 
$\gamma_{\eta} = 2\gamma_T$ and in Subsect. B 
$\gamma_{\eta} =-\gamma_{T}$.

By this point it should begin to appear evident that the differences in
the various cases emerge primarily in the exponent of the decay function.
This becomes fully clear once the basic equations are examined below.
In fact it can be recognized that the $\varphi_0$-equation of motion
Eq.(\ref{eom1}) actually is a equation of motion for $D(\tau)$. Using
Eq. (\ref{dissgen}) it has the general form
\begin{equation}
\frac{d\varphi_0(\tau)}{d\tau} =\varphi_{BI} \frac{dD(\tau)}{d\tau}
=-\frac{\pi^2\lambda \varphi^3_{BI}}{6 \eta'\Gamma' \kappa_M^2 Ng^2M^2}
D(\tau)^{3-\gamma_{\eta}} 
=-\frac{\pi^2\lambda \varphi_0^3(\tau)}{6 \eta'\Gamma' Ng^2 T^2(\tau)}
D(\tau)^{2\gamma_{T}-\gamma_{\eta}}.
\label{eomgen}
\end{equation}
On the other hand, from the general form of the $\varphi_0(\tau)$ solution
Eq.(\ref{varphigs}), it also follows that
\begin{equation}
\frac{d\varphi_0(\tau)}{d\tau} \propto 
\frac{y \varphi_0(\tau)}{(y\tau+1)^{1-\delta_e}},
\label{phialt}
\end{equation}
where $\delta_e=1$ if $D(\tau) \sim exp(-y\tau)$ and zero otherwise.
Based on this relation, we find that
$d \rho_{\phi}(\varphi_0)/d\tau = \lambda \varphi_0^3(\tau)
d\varphi_0(\tau)/d\tau/6 \propto 
y\rho_{\phi}(\varphi_0)/(y\tau+1)^{1-\delta_e}$
which applied to the energy conservation equation Eq.(\ref{dotr})
implies
\begin{equation}
\frac{\rho_{\phi}(\tau)}{\rho_r(\tau)}
\equiv \frac{2\lambda 
\varphi_0^4(\tau)}{3\pi^2[N_r+2\kappa_M ND^{\gamma_{\rho'}} (\tau)]
T^4(\tau)} \sim \frac{H(\tau)}{\Gamma_{\chi}(M)} \frac{y\tau+1}{y}
\sim \frac{N_e \beta}{h(\beta)} (y\tau+1)D^{\gamma_{\rho''}}(\tau),
\label{rhogen}
\end{equation}
where $\gamma_{\rho'}$ and $\gamma_{\rho''}$ are more exponents that
depend on the specific solution regime. In Subsects. A and B 
$\gamma_{\rho'}=\gamma_T$ and $0$ respectively.  $\gamma_{\rho''}$
represents the time dependence of $H$.  Thus $\gamma_{\rho''}=0$ in
all the cases except the last, IIIB2, where $\gamma_{\rho''}=2$.

Now we can deduce all the general features of the solutions in the
previous two subsections.  In addition to examining the solutions in terms
of the convenient parameters used in our analysis, it is interesting to
see how the solutions depend on the parameters with direct physical
interpretation, the Hubble parameter $H$ and the slope parameter for
$\varphi_0(\tau)$, $y$ in exchange of $\alpha$ and $N_e$.
{}For $\rho_{\phi}(\tau)/\rho_r(\tau)$, the general expressions already
have been given in Eq. (\ref{rhogen}).  {}From Eqs. (\ref{eomgen})
and (\ref{phialt}) we find
\begin{equation}
\frac{\lambda \varphi_0^2(\tau)}{2T^2(\tau)} \sim \eta' \Gamma' y N g^2
\frac{D^{\gamma_{\eta}-2\gamma_T}(\tau)}{(y\tau+1)^{1-\delta_e}}
= \frac{\eta' \Gamma' \alpha h(\beta) \kappa_M  N g^2}{N_e}
\frac{D^{\gamma_{\eta}-2\gamma_T}(\tau)}{(y\tau+1)^{1-\delta_e}}.
\label{mphigen}
\end{equation}
{}From Eqs. (\ref{rhogen}) and (\ref{mphigen}) we find
\begin{equation}
\lambda \sim (\eta' \Gamma')^2 \frac{y^3 N^2 g^4}
{(H/\Gamma)(N_r+2\kappa_MN)} =
(\eta' \Gamma')^2 \frac{\alpha^2 h^3(\beta) \kappa_M^2 N^2 g^4}
{\beta N_e^3 (N_r+2\kappa_MN)}
\label{lamgen}
\end{equation}
and
\begin{equation}
\frac{\varphi_{BI}}{M} \sim \frac{1}{(\eta'\Gamma')^{1/2}}
\left(\frac{(H/\Gamma) \kappa_M^2 (N_r+2\kappa_MN)}{y^2 Ng^2}
\right)^{1/2}
= \frac{1}{(\eta'\Gamma')^{1/2}}
\frac{\beta^{1/2} \kappa_M^{1/2} N_e (N_r+2\kappa_MN)^{1/2}}
{\alpha^{1/2} h(\beta) N^{1/2} g}.
\label{pomgen}
\end{equation}
Regarding the consistency conditions, the general forms are as follows:
$\varphi_0$-adiabatic condition Eq. (\ref{padiab})
\begin{equation}
\frac{y}{\kappa_M} 
\frac{D^{-\gamma_T}(\tau)}
{(y\tau+1)^{1-\delta_e}} =
\frac{\alpha h(\beta)}{N_e} 
\frac{D^{-\gamma_T}(\tau)}
{(y\tau+1)^{1-\delta_e}} < O(1),
\label{padaibgen}
\end{equation}
the thermal-adiabatic condition Eq. (\ref{thermadiab})
\begin{equation}
\left(\frac{(H/\Gamma)(N_r+2\kappa_M
N)}{\eta' \Gamma'
\kappa_M^2 N} \right)^{1/2}
\frac{D^{1-2\gamma_T}(\tau)}
{(y\tau+1)^{1-\delta_e}} =
\left(\frac{(\alpha \beta (N_r+2\kappa_M N)}
{(\eta' \Gamma') \kappa_M N}\right)^{1/2}
\frac{D^{1-2\gamma_T}(\tau)}
{(y\tau+1)^{1-\delta_e}} < O(1),
\label{tadaibgen}
\end{equation}
and the force condition Eq. (\ref{forcec})
\begin{eqnarray}
(\eta'\Gamma')^{1/2} 
\left(\frac{(H/\Gamma) (N_r+2\kappa_MN)}{N}\right)^{1/2} 
D^{3-3\gamma_T}(\tau) & \sim & 
(\eta'\Gamma')^{1/2} 
\left(\frac{\alpha \beta \kappa_M (N_r+2\kappa_MN)}{N}\right)^{1/2} 
D^{3-3\gamma_T}(\tau) \nonumber\\
& > & O(1).
\label{forcegen}
\end{eqnarray}

These expressions contain interesting insight into the solutions. 
The expression for $\rho_v/\rho_r \propto V(\varphi_0)/T^4$, 
Eq. (\ref{rhogen}), indicates that the relative content of radiation to vacuum
energy is directly proportional to the Hubble expansion rate $H_{BI}$
and inversely proportional to the slope parameter $y$.  These general
trends can be explained.  The former is expected since for a slower the
Hubble expansion rate, the red-shifting of radiation is slower. The
latter follows since as the slope parameter decreases, the decay of
vacuum energy to radiation also becomes slower.

The expression for $m_{\phi}^2/T^2$, Eq. (\ref{mphigen}), also has a simple
explanation. It is independent of the Hubble parameter, which is
indicative that this expression is entirely an outcome of the $\varphi_0$
equation of motion.  Since the potential is a monomial in $\varphi_0$ 
and the $\varphi_0$ equation of motion is first order and overdamped,
Eq. (\ref{phialt}) follows.  {}Furthermore, in our case the
$\varphi_0$ equation of
motion always has two powers of the temperature in the form
$d\varphi_0/d\tau \propto (dV(\varphi_0)/d\varphi_0)/T^2$.
Comparing this with $d\varphi_0/d \tau \sim y \varphi_0$ from
Eq. (\ref{phialt}) and noting that
our potential is quartic, $V(\varphi_0) \sim \lambda \varphi_0^4$, the
expression for $\lambda \varphi_0^2/T^2$ in Eq. (\ref{mphigen}) follows.

Based on these two expressions, Eqs. (\ref{rhogen}) and (\ref{mphigen}),
the remaining expressions follow. {}For example since 
$V(\varphi_0)/T^4 \propto H_{BI}/y$ 
and $(dV(\varphi_0)/d\varphi_0)/(\varphi_0 T^2) \propto y$, it
must follow that $\varphi_0^2/T^2 \propto H_{BI}/y^2$. It then implies
that for a quartic potential $\lambda \propto y^3/H_{BI}$. {}For a
hypothetical monomial potential $V(\varphi_0) \sim \lambda M^{4-n} \phi^n$, it
still follows that $\varphi_0^2/T^2 \propto H_{BI}/y^2$
but now $\lambda (\varphi_0/M)^{n-4} \propto y^3/H_{BI}$. The thermal
adiabatic condition, $\propto (d\varphi_0/d\tau)/T$, also can be
understood since 
$(d\varphi_0/d\tau)/T \propto y\varphi_0/T \propto H_{BI}^{1/2}$.
The behavior of the other expressions as well as other parameter
dependencies can be deduced by similar reasoning.

\section{Estimates for amplitude of density Perturbations
$\delta \rho/\rho$}
\label{sect5}

In this section estimates are made of scalar metric perturbations
for flat spatial geometry, $\Omega =1$, for the four cases in the
previous section.  For the calculations to follow, 
the basic formulas can be deduced either
from first principles arguments or
from a perturbative quantum field theory calculation,
and we will take the former route.
This approach has limitations, within which it
is indisputable.  For our purposes these limitations
are unimportant in obtaining the basic formulas and in making
the estimates 
for density perturbations in the quantum
field theory model studied in this paper.
The limitation of this approach is in sacrificing  
some of the details
that could be obtained by the perturbative calculation.
These details, which will be described at an appropriate
place below,  are not useful for our present purposes. Nevertheless,
at some stage, it will be important to understand these details and
to do the perturbative calculation. For this purpose, the
present approach also is useful as a cross-check
for the perturbative calculations and in providing
a necessary outline
of the formal problems that must be addressed
for that derivation.

To estimate density perturbations, the classical background field
now is treated with fluctuations, 
$\varphi({\bf x},t) = \varphi_0(t)+ \delta \varphi({\bf x},t)$,
where $\varphi_0(t)$ is the zero mode and 
$\delta \varphi({\bf x},t)$ are small fluctuations about the zero mode.
The calculations to follow are in momentum space. The Fourier transform
of the fluctuations is defined as
\begin{equation}
\delta \varphi({\bf k},t) = \int_{V} d^3x 
\delta \varphi({\bf x},t) \exp(-i {\bf k} \cdot {\bf x})
\end{equation}
where $({\bf k},{\bf x})$ can be either comoving coordinates or
physical coordinates at a particular time.  Hereafter 
we denote comoving coordinates as
$({\bf k}_c,{\bf x}_c)$ and physical coordinates at time $\tau$ as
$({\bf k}_p(\tau),{\bf x}_p(\tau))$, where the $\tau$-
dependence sometimes is not shown explicitly. 
The comoving coordinates can be regarded as
the intrinsic labels for the modes of the scalar field fluctuations.
Thus the modes will be denoted as 
$\delta \varphi_{{\bf k}_c}(\tau)$.  {}For definiteness, the comoving and
physical coordinates will coincide at the end of warm inflation
$\tau_{EI}$ so that for any
component direction $k_p(\tau) = k_c/R(\tau-\tau_{EI})$,
where by our convention $R(0)=1$.  {}For the four cases treated in the
last section $R(\tau) \approx \exp[H(\tau) \tau]$.
Very often it will be necessary to consider modes in terms of their
physical wavenumber at a given time $\tau$. Thus, the modes also may be
denoted as $\delta \varphi({\bf k}_p,\tau)$, where 
the (redundant) ${\bf k}_c$ subscript has been dropped.

Let us obtain the equation of
motion for the fluctuations in flat nonexpanding spacetime.
We argue below that this equation also suffices for our purposes in
an expanding background.  {}For the nonexpanding background case,
no distinction is necessary between comoving and physical
wavenumbers; the modes are
denoted simply as $\delta \varphi({\bf k},\tau)$.
 
As stated earlier, the equations of motion for the
fluctuations can be obtained by two approaches, either by general
first principles arguments or a direct perturbative calculation,
and here the former approach is taken.  The basic fact that permits
the former approach is that the calculation of the zero mode
and fluctuation equations of motion, which follow
the methods of \cite{hs1,morikawa,lawrie,gr,bgr},
are done in a near-thermal-equilibrium approximation, thus 
respecting the fluctuation-dissipation theorem.  This fact
immediately determines the form of the fluctuation equation
of motion, given that for the zero mode.
In particular, assuming the fluctuations are small,
their equation of motion is obtained through a linearization
of the zero mode equation of motion, and with the inclusion
of the gradient term, $\nabla^2 \delta \varphi$, and a noise term
that represents the short distance dynamics of the heat bath.
The fluctuation dissipation theorem uniquely determines the correlation
statistics of the noise.  The missing detail from this
first principles procedure, which were eluded to earlier,
is the relation of the noise function to the basic field
variables of the Lagrangian.  For our present purposes,
this relation is not needed.  Examples of such relations obtained
from perturbative calculations are given in \cite{morikawa,gr}

Turning to the equation of motion for the fluctuations,
as stated above,  it is obtained through the linearized
deviation of the zero mode equation of motion,
with account for the gradient term and noise.  Although
our interest in the zero mode dynamics is in the
overdamped regime, where its equation of motion becomes
first order in time, for obtaining the fluctuation
equation of motion we must start with the initial second order
zero mode equation.  Compared to the zero mode equation
of motion, the only additional term in the fluctuation equation
of motion that could make it second order is the gradient term.
This will happen if the gradient term creates a sufficiently
large curvature ($\equiv -\nabla^2 \phi + m_{\phi}^2 \phi$)
in the potential to overcome the large damping 
force term.  However, this will not happen since our interest
is in fluctuations that relatively are not very
large.  In particular, we will see below from explicit
calculations that for the fluctuation wavenumbers of interest,
the curvature term ($=\sqrt{{\bf k}^2+m_{\phi}^2}$)
generally is smaller than the dissipative coefficient
(recall this is the term multiplying $d\varphi_0/dt$ in
Eq. (\ref{eom1})), and generically for a damped
harmonic oscillator equation of motion this implies the
overdamped (first order) regime. 
Thus, the equation of motion for the fluctuation
$\delta \varphi({\bf k},\tau)$ is obtained from the equation of
motion for $\varphi_0(\tau)$ that is given in Subsect. \ref{subsect3A}
Eq. (\ref{eom1})
and derived in \cite{bgr},  
\begin{equation}
\Upsilon(\varphi_0,T) 
\frac{d \delta \varphi({\bf k},\tau)}{d \tau} = 
-({\bf k}^2+m^2_{\phi}(\varphi,T)) \delta \varphi({\bf k},\tau)
+\xi({\bf k},\tau)
\label{eomfluc}
\end{equation}
where
\begin{equation}
\Upsilon(\varphi_0,T)  \equiv 
\sum_i^{t.e.}(\varphi_0-M_i)^2\eta_1(T) \Gamma_{\chi}(M)
\label{upsdef}
\end{equation}
and $m_{\phi}(\varphi_0,T)$ is given in Eq. (\ref{rmphi}).
This equation is obtained from Eq. ({\ref{eom1}) by substituting
$\varphi({\bf x},t)$ and retaining terms linear in 
$\delta \varphi({\bf x},t)$ and Fourier transforming
to k-space. In addition a noise function
$\xi({\bf k}_p,t)$ is added which respects the fluctuation-dissipation
theorem \cite{fdt}
\begin{equation}
<\xi({\bf k},t)>_{\xi} =0
\end{equation}
\begin{equation}
<\xi({\bf k},\tau) \xi(-{\bf k}^{\prime},\tau^{\prime})
>_{\xi} \stackrel{T\rightarrow \infty}{=}
2 \Upsilon(\varphi_0(\tau),T(\tau)) T(\tau)
(2\pi)^3 \delta^{(3)}(\bf{k}-\bf{k}) \delta(\tau-\tau^{\prime}).
\end{equation}
The equation of motion (\ref{eomfluc}) represents the standard
near-thermal-equilibrium dynamics in which $\xi({\bf k},\tau)$ drives the
correlations of $\delta \varphi({\bf k},\tau)$  to thermal equilibrium
with the relaxation rate of the initial conditions 
$({\bf k}^2+m^2_{\phi}(\varphi_0,T))/\Upsilon(\varphi,T)$.

The relevance of Eq. (\ref{eomfluc}) to expanding background is that
it approximates the equation of motion for a mode $\delta
\varphi_{{\bf k}_c}(\tau)$ during a Hubble time interval say
$\tau_i \stackrel{<}{\sim} \tau \stackrel{<}{\sim} \tau_{i+1}$ with 
$\tau_{i+1} - \tau_i \approx \Gamma_{\chi}(M)/H$.  {}Furthermore, the momentum
vector in the equation of motion Eq. (\ref{eomfluc}) is identified with the
physical momentum of ${\bf k}_c$ which to a good approximation
is fixed at one intermediate
time during the respective time interval, such as $\tau_i$,
${\bf k} \rightarrow {\bf k}_p = {\bf k}_c \exp[-H(\tau)(\tau_i-\tau_{EI}]$.
This approximation is valid for large ${\bf k}_p$, when
the evolution of
$H(\tau)$,${\bf k}_p(\tau)$ and $\varphi(\tau)$ is
adiabatic relative to the evolution of $\delta \varphi_{{\bf k}_c}(\tau)$
during the respective time interval.
Within the regime where this approximation is valid, the evolution of
$\delta \varphi_{{\bf k}_c}(\tau)$ can be computed through piecewise
construction of solutions for a sequence of Hubble time intervals,
similar to the demonstration in \cite{ab1}.
We will see below that the regime of ${\bf k}_p$ where the above
approximation holds also is the appropriate regime for our purposes
of estimating density perturbations.

Consider what happens to a mode $\delta \varphi_{{\bf k}_c}(\tau)$
that is immersed in a heat bath and is in an expanding background.  The
larger ${\bf k}^2_p$ is, the faster is the relaxation rate.
If ${\bf k}_p^2$ is sufficiently large for the mode to relax 
within a Hubble time, then that mode thermalizes.
Thus at any instant during expansion, one can expect modes
with physical momenta bigger than some lower bound $k_F$
to thermalize within a Hubble time interval.  {}For these modes,
within a single Hubble time interval, the flat-space equation
of motion for the fluctuations Eq.(\ref{eomfluc}) is approximately valid.

As soon as the physical wavenumber of a $\varphi({\bf x},\tau)$ field
mode becomes less than $k_F$, it essentially feels no effect of the
thermal noise $\xi({\bf k}_p,\tau)$ during a Hubble time.  Thus for
mode $\delta \varphi_{{\bf k}_c}(\tau)$, it essentially does not change
once 
$|{\bf k}_p| \equiv |{\bf k}_c|\exp[-(H(\tau)/\Gamma_{\chi}(M))(\tau-\tau_{EI})]
<k_F$, and at $|{\bf k}_p| =k_F$ the mode assumes its thermalized
distribution.  If $k_F > H(\tau)$, it implies the $\delta \varphi_{{\bf
k}_c}(\tau)$ correlations that must be computed at time of Hubble radius
crossing, $|{\bf k}_p(\tau)|=H(\tau)$, are the thermalized correlations that
were fixed at $|{\bf k}_p(\tau)|=k_F$.  This effect was clarified 
for warm inflation by Yokoyama
and Linde \cite{yl} and they referred to it as ''freeze-out''.

In order to determine $k_F$, consider the solution of Eq.
(\ref{eomfluc}) for $\delta \varphi_{{\bf k}_c}(\tau)$ within the Hubble
time interval $\tau_0 < \tau < \tau_0+\Gamma/H(\tau_0)$.  
We will ignore the time variation of 
${\bf k}_p$,
$\Upsilon(\varphi_0,T)$, and $m_{\phi}(\varphi_0,T)$
during this time interval.  Their values at $\tau_0$ will be used
${\bf k}_p(\tau_0)$,
$\Upsilon(\tau_0) \equiv \Upsilon(\varphi_0(\tau_0,T(\tau_o))$, 
and $m_{\phi}(\tau_0) \equiv m_{\phi}(\varphi_0(\tau_0),T(\tau_0))$.
The approximate solution then is
\begin{eqnarray}
\delta \varphi_{{\bf k}_c}(\tau) & \approx & \frac{1}{\Upsilon(\tau_0)}
\exp[-\frac{{\bf k}_p^2+m^2_{\phi}(\tau_0)}{\Upsilon(
\tau_0)}(\tau-\tau_0)]
\int_{\tau_0}^{\tau}
\exp[\frac{{\bf k}_p^2+m^2_{\phi}(\tau_0)}{\Upsilon(
\tau_0)}(\tau'-\tau_0)]
\xi({\bf k}_p,\tau') d\tau' \nonumber\\
& + &
\delta \varphi_{{\bf k}_c}(\tau_0)
\exp[-\frac{{\bf k}_p^2+m^2_{\phi}(\tau_0)}{\Upsilon(
\tau_0)}(\tau-\tau_0)].
\label{phiksol}
\end{eqnarray}
and for the corresponding correlation function
\begin{eqnarray}
\langle \delta \varphi_{{\bf k}_c}(\tau) \delta \varphi_{{\bf k}'_c}(\tau)
\rangle_{\xi} & \approx &
(2\pi)^3 \delta^{(3)}({\bf k}_p - {\bf k}_p')
\frac{T}{[{\bf k}_p^2+m^2_{\phi}(\tau_0)]}
\left[1.0-\exp\left(-\frac{2({\bf k}_p^2+m^2_{\phi}(\tau_0))}
{\Upsilon(\tau_0)}
(\tau-\tau_0)\right)\right] \nonumber\\
& + &
\langle \delta \varphi_{{\bf k}_c}(\tau_0) \delta \varphi_{{\bf k}'_c}(\tau_0)
\rangle_{\xi}
\exp\left[-\frac{2({\bf k}_p^2+m^2_{\phi}(\tau_0))}{\Upsilon(\tau_0)}
(\tau - \tau_0) \right].
\end{eqnarray}
When the exponentially decaying terms are negligible, the above
correlation is equivalent to the high-temperature correlation function 
$\langle \delta \varphi({\bf k}_p,\tau_0) \delta \varphi({\bf k}_p',\tau_0)
\rangle_{T}$.

In this solution Eq. (\ref{phiksol}), on the right hand side, 
the first term is the noise
contribution that is driving the mode to thermal equilibrium and the
second term contains the memory of the initial conditions at
$\tau=\tau_0$, which are exponentially damping.
By definition of freeze-out, for 
$|{\bf k}_p| \stackrel{>}{\sim} k_F$ the second term damps
away within a Hubble time and for $|{\bf k}_p| \stackrel{<}{\sim} k_F$ 
it does not. To quantify
the criteria, the freeze-out momentum $k_F$ is defined by the condition
\begin{equation}
\frac{{\bf k}_F^2+m^2_{\phi}(\varphi(\tau),T(\tau))}{\Upsilon(
\varphi_0(\tau),T(\tau))} \frac{\Gamma_{\chi}(M)}{H(\tau)} = 1.
\label{kfcond}
\end{equation}
This relation allows us to
follow-up our discussion above Eq. (\ref{eomfluc}) 
about the fluctuation equation
of motion being first order in time.  From the above equation
we see that the curvature term in the potential 
$\sqrt{{\bf k}^2 + m_{\phi}^2}$ is less than the dissipative coefficient
(in this notation comparing to Eq. (\ref{eom1}) 
the dissipative coefficient is $\Upsilon/\Gamma_{\chi}$)
since generically $H < \Upsilon/\Gamma_{\chi}$. As such, the overdamped
equation of motion for the fluctuations, 
Eq. (\ref{eomfluc}), is justified for the 
wavenumbers of interest to us. 

Now we can write down the basic equations for calculating density
perturbations during warm inflation.  Properly this should be fully
derived from the linearized General Relativity equations 
for perturbations
\cite{bardeen,sasaki2,brandpert}. 
Nevertheless, an examination of the adiabatic
density perturbations derivation for supercooled inflation \cite{guthpi1,amp}
with the modification of a subdominant radiation
component for the warm inflation case leaves unaltered the
basic expression for $\delta \rho/\rho$. 
This conclusion was arrived at
in our earlier papers \cite{bf2,wi} and independently
by Yokoyama and Linde \cite{yl}.
Therefore from this we find
\begin{equation}
\frac{\delta \rho}{\rho}(\tau)
= \frac{V'(\varphi_0) \Delta \varphi_{H(\tau)}[k_F(\tau-{\tilde
\tau}(\tau))]}{(\Gamma_{\chi}(M)d\varphi_0/d\tau)^2 +(4/3)\rho_r(\tau)}
\approx \frac{6 H(\tau) \Delta \varphi_{H(\tau)}[k_F(\tau-{\tilde
\tau}(\tau))]}{5 \Gamma_{\chi}(M)(d\varphi_0/d\tau)}.
\label{delrho}
\end{equation}
The middle expression is the one used by Berera and Fang
\cite{bf2} and the latter is the Guth and Pi expression \cite{guthpi1}.
During warm inflation, since 
from Eq. (\ref{dotr}) $\rho_r \approx V'({\varphi_0}) 
\Gamma_{\chi}(M) (d\varphi_0/d\tau)/(4H(\tau))$ and 
$\rho_r \gg (\Gamma_{\chi}(M) d\varphi_0/d\tau)^2$,
the middle and left expressions above are equivalent up to an
O(1) constant.  It
would be interesting to investigate which of the two expressions is
fundamentally more proper.  In Eq. (\ref{delrho}),
$\Delta \varphi_{H(\tau)}[k_F(\tau-{\tilde \tau}(\tau)]$ is the
amplitude of the scalar field fluctuations and it is composed of all the
modes whose physical wavelengths cross the Hubble 
radius within one Hubble time interval about
time $\tau$.   Recall that the comoving mode 
of physical wavenumber $|{\bf k}_p (\tau)| \sim
H(\tau)$ had its amplitude frozen at an earlier time $\tau'$ when its
physical wavenumber was $|{\bf k}_p (\tau')| \sim k_F(\tau')$. Since
$|{\bf k}_p(\tau')|/|{\bf k}_p(\tau)| =\exp[(H(\tau) \tau - H(\tau')
\tau')/\Gamma_{\chi}(M)]$, this implies 
$k_F(\tau') =H(\tau) \exp[(H(\tau)\tau - H(\tau')
\tau')/\Gamma_{\chi}(M)]$.  This defines
${\tilde \tau}(\tau) \equiv \tau - \tau'$.
{}For the cases in the previous sections, $k_F(\tau)$ is slowly varying,
so we will use the approximation 
$k_F(\tau-{\tilde \tau}(\tau)) \approx k_F(\tau)$.
The expression for the scalar field amplitude is defined as the natural
finite-temperature
extension of the $T=0$ expression of supercooled
scalar field inflation and with account for $k_F$.  
We use the definition 
\begin{equation}
\Delta \varphi_H^2[k_F] \equiv
\int_{k_F-shell}\frac{d^3{\bf k}_p}{(2\pi)^3} 
\int_{V} d^3{\bf x}_p 
\langle \delta \varphi({\bf x},\tau)
\delta \varphi({\bf 0},\tau) \rangle_T
\exp(-i{\bf k}_p \cdot {\bf x}_p)
\stackrel{T \rightarrow \infty}{\approx}
\frac{k_F T}{2 \pi^2},
\label{Dvp}
\end{equation}
where the $k_F$ - shell is defined as the spherical shell which is bounded
between $k_F {\rm e}^{-1/2} < |{\bf k}_p| < k_F {\rm e}^{1/2}$
(we approximate the shell thickness simply to be $k_F$).
The expression on the far right in the above equation is valid when
$k_F < T$.
The definition Eq. (\ref{Dvp})
is equivalent to the one given by Linde
\cite{lindebook} in which one retains from
$<\phi^2({\bf x}, \tau)>_T$ the contribution from wavenumbers within the
$k_F$-shell.  {}From Eq.(7.3.2) and Eq. (3.1.7) of Linde's
book \cite{lindebook}
this gives
\begin{equation}
\Delta \varphi_H^2[k_F] \equiv
\frac{1}{2\pi^2} \int_{k_F-shell}
\frac{k^2 dk}{\sqrt{k^2+m_{\phi}}
\left[\exp(\sqrt{k^2+m_{\phi}}/T) -1 \right]}. 
\label{Dvplb} 
\end{equation}

When $k_F > T$, Eq. (\ref{Dvp}) implies 
$\Delta \varphi_H^2(k_F)_{k_F > T} = k_F^2/(4\pi^2)$.
{}For this region, it is a poor approximation to use the zero-mode dissipative
coefficient in the k-mode equation of motion Eq. (\ref{eomfluc}).
{}For this regime, a proper calculation of the k-mode
dissipative function is important.  Our calculations in the next two
subsections consider only the high temperature expression
on the right hand side of Eq. (\ref{Dvp}).

Substituting the expression for $\Delta \varphi^2_H$ from
Eq. (\ref{Dvp}) into Eq. (\ref{delrho}), the final expression for
$\delta \rho/\rho$ for the mode that crosses the Hubble radius at
time $\tau$ during warm inflation
is\footnote{In our estimates of density perturbations, 
the equation of motion for
$\delta \varphi_{{\bf k}_c}(\tau)$, Eq. (\ref{eomfluc}) uses the
${\bf k}=0$ dissipative function Eq. (\ref{eta1B}).
The derivation  in \cite{gr,bgr}
suggests that the dissipative function
$\eta_1(T)$ decreases as ${\bf k}$ increases, since it requires 
``off-diagonal'' Green's functions.  This would imply $k_F$ decreases
since the relaxation rate is faster, thus any mode 
$\delta \varphi_{{\bf k}_c}(\tau)$ would thermalize faster at every
physical wavenumber.  Ultimately this means for a given ${\bf k}_c$ mode
that is crossing the Hubble radius, $\delta \rho/\rho$ decreases relative
to our estimates.
In the future, a proper derivation of
the dissipative function at non-zero k-vector would be useful.}  
\begin{equation}
\frac{\delta \rho}{\rho}(\tau) \stackrel{k_F < T}{\approx}
\frac{6[H(\tau)/\Gamma_{\chi}(M)] \sqrt{k_F(\tau) T(\tau)}}
{5\sqrt{2} \pi d\varphi_0(\tau)/d\tau}.
\label{delrhof}
\end{equation}
Recall, the comoving mode that crosses the Hubble radius at time
$\tau$ is $|{\bf k}_c| \approx H(\tau) \exp[(H(\tau)/\Gamma_{\chi}(M))
(\tau-\tau_{EI})]$.

{}From the above considerations, the final prescription for
computing $\delta \rho/\rho$ is simple.  {}First determine
the freeze-out wavenumber from Eq. (\ref{kfcond}),
and then substitute this in
Eq.(\ref{delrhof}) along with expressions for $T(\tau)$, $H(\tau)$ and
$d\varphi_0(\tau)/d\tau$ from the previous section.
In the next two subsections this is done for the cases in the
coinciding subsections of the last section.  {}Finally
Subsect. \ref{subsect5C} follows with a discussion of the results.
 
\subsection{$T_{BI} \gg T_{EI}$}
\label{subsect5A}

This subsection considers the cases in Subsect. A of the last
section.

\subsubsection{\bf $N_r=0$}
\label{subsubsect5A1}

{}From Eq. (\ref{eomfluc}) and taking $\Upsilon$ in Eq. (\ref{upsdef})
the same as
in  Eq. (\ref{eomA}), the equation of motion for
the fluctuation is
\begin{equation}
\frac{d \delta \varphi({\bf k}_p,\tau)}{d\tau} =
-[{\bf k}^2_p + m_{\phi}^2(\varphi,T)] \frac{\pi^2}{\eta'\Gamma'ng^2
T^2(\tau)} \delta \varphi({\bf k}_p,\tau)
=-21 y \left(\frac{\varphi_0(\tau)}{\varphi_{BI}}\right)^{2/7}
\frac{{\bf k}^2_p}{\lambda \varphi_0^2(\tau)}
\delta \varphi({\bf k}_p,\tau).
\end{equation}
The expression on the extreme right was obtained by substituting
the expression for $y$ below Eq. (\ref{phia1}) and considering the
regime where ${\bf k}_p^2 \gg m_{\phi}^2(\varphi_0,T)
\approx \lambda \varphi_0^2(\tau)/2$.  The freeze-out wavenumber from
Eq. (\ref{kfcond}) is
\begin{equation}
k^2_F(\tau) = \frac{(H_{BI}/\Gamma_{\chi}(M)) \lambda \varphi_0^2(\tau)}{21y}
\left(\frac{\varphi_0(\tau)}{\varphi_{BI}}\right)^{2/7}.
\end{equation}
Using this, from Eq.(\ref{delrhof}) we find
\begin{eqnarray}
\frac{\delta \rho}{\rho}(\tau) & \approx &
\frac{6 (\eta'\Gamma')^{3/4}}{5 \pi^{7/2}}
\left(\frac{H_{BI}}{\Gamma_{\chi}(M)}\right)^{3/4}
\kappa_M^{-1/2} N^{1/4} g^{3/2} (y\tau+1)^{5/4} \nonumber\\
& = & \frac{6 (\eta'\Gamma')^{3/4}}{5 \pi^{7/2}}
\alpha^{3/4} \beta^{3/4} \kappa_M^{1/4} N^{1/4} g^{3/2}
(y\tau+1)^{5/4}.
\end{eqnarray}
The spectrum is very flat. Since the relation between time $\tau$
and the comoving wavenumber ${\bf k}_c$ crossing the Hubble radius at
time $\tau$ is 
$\tau = -(\Gamma_{\chi}(M)/H(\tau))\ln(|{\bf k}_c|/H_{EI}) + \tau_{EI}
\approx -(\Gamma_{\chi}(M)/H_{BI})\ln(|{\bf k}_c|/H_{BI}) + \tau_{EI}$, 
the variation of
$\delta \rho/\rho \propto 
[y(-(\Gamma_{\chi}(M)/H_{BI})\ln(|{\bf k}_c|/H_{BI})
+\tau_{EI})+1]^{5/4}$
is only logarithmic. {}For example,
for arbitrary e-folds $N_e$, $\delta \rho/\rho$ 
for the last scale that crosses the Hubble radius
(smallest scale) is a factor $\beta^{-5/12}$  bigger than the first scale
that crosses the Hubble radius (largest scale).
Observe that this deviation from exact scale invariance is
an outcome of the nontrivial thermodynamics and is not, in particular,
initially inputed by hand by choosing a nonanalytic potential.

\subsubsection{\bf $N_r \gg 2NT/T_{EI}$}
\label{subsubsect5A2}

{}From Eq. (\ref{eomfluc}) and taking $\Upsilon$ in Eq. (\ref{upsdef})
the same as
in  Eq. (\ref{eomA}), the equation of motion for
the fluctuation is
\begin{equation}
\frac{d \delta \varphi({\bf k}_p,\tau)}{d\tau} =
-[{\bf k}^2_p + m_{\phi}^2(\varphi,T)] \frac{\pi^2}{\eta'\Gamma'Ng^2
T^2(\tau)} \delta \varphi({\bf k}_p,\tau)
=-6 y
\frac{{\bf k}^2_p}{\lambda \varphi_0^2(\tau)}
\delta \varphi({\bf k}_p,\tau).
\end{equation}
The expression on the extreme right was obtained by substituting
the expression for $y$ below Eq. (\ref{tempA2}) and considering the
regime where ${\bf k}_p^2 \gg m_{\phi}^2(\varphi_0,T)
\approx \lambda \varphi_0^2(\tau)/2$.  The freeze-out wavenumber from
Eq. (\ref{kfcond}) is
\begin{equation}
k^2_F(\tau) = \frac{(H_{BI}/\Gamma_{\chi}(M)) \lambda \varphi_0^2(\tau)}{6y}.
\end{equation}
Using this and from Eq.(\ref{delrhof}) we find
\begin{equation}
\frac{\delta \rho}{\rho}(\tau) \approx
\frac{6 \sqrt{2} (\eta'\Gamma')^{3/4}}{\pi^{7/2}}
\frac{(H_{BI}/\Gamma_{\chi}(M))^{3/4} N^{3/4} g^{3/2}}{N_r^{1/2}}
=\frac{6 \sqrt{2} (\eta'\Gamma')^{3/4}}{\pi^{7/2}}
\frac{\alpha^{3/4} \beta^{3/4} \kappa_M^{3/4} N^{3/4} g^{3/2}}
{N_r^{1/2}}.
\end{equation}
{}For this case, the spectrum is exactly flat, i.e. $\delta \rho/\rho$
is the same for all e-folds.

\subsection{$T_{BI} \stackrel{>}{\sim} T_{EI}$}
\label{subsect5B}

This subsection considers the cases in Subsect. B of the last
section.

\subsubsection{$V_0> 0$}
\label{subsubsect5B1}

{}From Eq. (\ref{eomfluc}) and taking $\Upsilon$ in Eq. (\ref{upsdef}), 
the same as
in  Eq. (\ref{eomB}), the equation of motion for
the fluctuation is
\begin{equation}
\frac{d \delta \varphi({\bf k}_p,\tau)}{d\tau} =
-[{\bf k}^2_p + m_{\phi}^2(\varphi,T)] \frac{\pi^2 T(\tau)}{\eta'\Gamma'Ng^2
\kappa_M^3 M^3} \delta \varphi({\bf k}_p,\tau)
=-\frac{3}{2}y \left(\frac{\varphi_0(\tau)}{\varphi_{BI}}\right)^4
\frac{{\bf k}^2_p}{\lambda \varphi_0^2(\tau)}
\delta \varphi({\bf k}_p,\tau).
\label{eomflucvog0}
\end{equation}
The expression on the extreme right was obtained by substituting
the expression for $y$ below Eq. (\ref{tempB1}) and considering the
regime where ${\bf k}_p^2 \gg m_{\phi}^2(\varphi_0,T)
\approx \lambda \varphi^2_0(\tau)/2$.  The freeze-out wavenumber from
Eq. (\ref{kfcond}) is
\begin{equation}
k^2_F(\tau) = \frac{2(H(\tau)/\Gamma_{\chi}(M)) \lambda \varphi_0^2(\tau)}{3y}
\left(\frac{\varphi_{BI}}{\varphi_0(\tau)}\right)^4.
\end{equation}
Using this and from Eq. (\ref{delrhof}) we find
\begin{eqnarray}
\frac{\delta \rho}{\rho}(\tau) & \approx &
\frac{6 \sqrt{2} (\eta'\Gamma')^{3/4}}{5\pi^{7/2}}
\frac{(H_{BI}/\Gamma_{\chi}(M))^{3/4} N^{3/4} g^{3/2}}{(N_r+2\kappa_MN)^{1/2}}
(y\tau+1)^{9/8} \nonumber\\
& = &\frac{6 \sqrt{2} (\eta'\Gamma')^{3/4}}{5 \pi^{7/2}}
\frac{\alpha^{3/4} \beta^{3/4} \kappa_M^{3/4} N^{3/4} g^{3/2}}
{(N_r+2\kappa_M N)^{1/2}}(y\tau+1)^{9/8}.
\end{eqnarray}
{}The spectrum is nearly flat with deviations that are
logarithmic.  {}For arbitrary e-folds $N_e$, $\delta \rho/\rho$ at the last
e-fold (smallest scale) is only a factor $\beta^{-9/4}$ bigger than at
the first e-fold (largest scale).  Recall that the regime for
this approximation requires $1 > \beta \stackrel{>}{\sim} 0.5$.

\subsubsection{$V_0 = 0$}
\label{subsubsect5B2}

{}From Eq. (\ref{eomfluc}) and taking $\Upsilon$ in Eq. (\ref{upsdef}) 
the same as
in  Eq. (\ref{eomB}), the equation of motion for
the fluctuation is
\begin{equation}
\frac{d \delta \varphi({\bf k}_p,\tau)}{d\tau} =
-[{\bf k}^2_p + m_{\phi}^2(\varphi,T)] \frac{\pi^2 T(\tau)}{\eta'\Gamma'Ng^2
\kappa_M^3 M^3} \delta \varphi({\bf k}_p,\tau)
=-\frac{9}{5}y \left(\frac{\varphi_0(\tau)}{\varphi_{BI}}\right)^{10/3}
\frac{{\bf k}^2_p}{\lambda \varphi_0^2(\tau)}
\delta \varphi({\bf k}_p,\tau).
\end{equation}
The expression on the extreme right was obtained by substituting
the expression for $y$ below Eq. (\ref{tempB2}) and considering the
regime where ${\bf k}_p^2 \gg m_{\phi}^2(\varphi_0,T)
\approx \lambda \varphi_0^2(\tau)/2$.  The freeze-out wavenumber from
Eq. (\ref{kfcond}) is
\begin{equation}
k^2_F(\tau) = \frac{5(H(\tau)/\Gamma_{\chi}(M)) \lambda \varphi_0^2(\tau)}{9y}
\left(\frac{\varphi_{BI}}{\varphi_0(\tau)}\right)^{10/3}.
\label{kfB2}
\end{equation}
In this case the Hubble parameter from Eq. (\ref{hubve0})
is $\tau$-dependent.  Using this expression and the above condition for 
$k_F$, from Eq. (\ref{delrhof}) we find
\begin{eqnarray}
\frac{\delta \rho}{\rho}(\tau) & \approx &
\frac{6 \sqrt{2} (\eta'\Gamma')^{3/4}}{5\pi^{7/2}}
\frac{(H_{BI}/\Gamma_{\chi}(M))^{3/4} N^{3/4} g^{3/2}}{(N_r+2\kappa_MN)^{1/2}}
(y\tau+1)^{9/20} \nonumber\\
& = &\frac{6 \sqrt{2} (\eta'\Gamma')^{3/4}}{5 \pi^{7/2}}
\frac{\alpha^{3/4} \beta^{3/4} \kappa_M^{3/4} N^{3/4} g^{3/2}}
{(N_r+2\kappa_M N)^{1/2}}(y\tau+1)^{9/20}.
\label{delrB2}
\end{eqnarray}
{}The spectrum is nearly flat with deviations that are
logarithmic.  {}For arbitrary e-folds $N_e$, $\delta \rho/\rho$ at the last
e-fold (smallest scale) is only a factor $\beta^{-9/8}$ bigger than at
the first e-fold (largest scale).  Again recall that the regime for
this approximation requires $1 > \beta \stackrel{>}{\sim} 0.5$.

\subsection{Discussion}
\label{subsect5C}

Some aspects of the results for density perturbations are discussed
here.  Recall from observation that $\delta \rho/\rho \sim
10^{-5}$ \cite{smoot}.  {}From our results, for all four cases we find
$\delta \rho/\rho \stackrel{<}{\sim} 10^{-2} \alpha^{3/4} \kappa_M^{3/4} N^{1/4}
g^{3/2}$.  Thus the parameters $\alpha$, $\kappa_M$, $N$ and $g$
must decrease $\delta \rho/\rho$ by at least three orders of magnitude.
{}For observational consistency it is permissible if 
$\delta \rho/\rho < 10^{-5}$ from inflation, since post-inflationary 
mechanisms such as cosmic strings also can produce density perturbations.

Overall consistency with respect to quantum field theory from Subsect. 
\ref{subsect3B}
and observation, 
$\delta \rho/\rho \stackrel{<}{\sim} 10^{-5}$ and 
$N_e \stackrel{>}{\sim} 60$, can 
be achieved in a variety of ways.  {}From the 
discussion Subsect. \ref{subsect4C},
Eq. (\ref{genalp}) is a general  consistency regime for quantum field
theory for arbitrary $N_e$.  Combining this with the general
form for $\delta \rho/\rho$ stated above, overall consistency can be
achieved for all four solution regimes in general for
\begin{equation}
\frac{1}{\beta \kappa_M^2} \stackrel{<}{\sim} \alpha \stackrel{<}{\sim}
{\rm min} \left(\frac{1}{\kappa_M^{1/3}},
\frac{\beta \kappa_M N}{100(N_r+2\kappa_M N)}\right).
\label{genalpd}
\end{equation}
Alternatively, for arbitrary $\alpha$,$\beta$, and $\kappa_M$,
$\delta \rho/\rho$ can be made arbitrarily small by requiring
$g \rightarrow 0$, $Ng^a = {\rm const.}$ with
$a<6$.  Thus the model provides sufficient freedom to achieve consistency
independently
with respect to quantum field theory and observation.
{}Furthermore, both alternatives for decreasing
$\delta \rho/\rho$, $\alpha \rightarrow 0$  or $g \rightarrow 0$ 
improve the validity of the underlying approximations. In the former
case the adiabatic regime is deepened and in the latter case
perturbation theory is further justified.

It is important to note the magnitude of the Hubble parameter within the
consistency regime.  {}From Eq.(\ref{hubdef}) recall that 
$H_{BI} = \alpha \beta \kappa_M \Gamma_{\chi}(M) 
\approx [\alpha \beta \kappa_M/(192 \pi)]
(g^4 +f^2/8)$, with thermalization condition Eq. (\ref{therm})
requiring $\alpha < 1$.  None of the consistency conditions so far have
imposed any conditions on the self-coupling parameter $f$ except
perturbation theory which requires 
$f \stackrel{<}{\sim} 1$.  Thus irrespective of $g$,
the bound can be given $H_{BI} \leq (2 \times 10^{-4}) \alpha \beta \kappa_M
M$.  

There is some freedom to this bound.
Observe that the entire calculation is independent of the number of
bosonic decay channels for the respective $\chi_{ik}$-field.  This reflects
itself in the coupling constant dependent factor 
${\cal C}$ in $\Gamma_{\chi}(T)$.
Note that in the basic equations the product
$\eta_1(T) \Gamma_{\chi}(M) $ always arises together, and because of this the
coupling constant dependent factor, in our model 
${\cal C} \approx (g^4+f^2/8)$, 
cancel. This is not a coincidence, since fundamentally 
$\eta_1(T) \sim 1/{\Gamma}$ as evident from the formalism of
\cite{gr,bgr} and from heuristic arguments in
\cite{hs1,yl}.  {}For the model in this paper, each $\chi_{ik}$ field has
only two decay channels, one to itself and the other to the $\phi$
field.  The $\chi_{ik}$ field also may interact with other
fields. {}For example consider the additional interaction
$(g^2_{\sigma}/2) \sum_{j=1}^{N_{\sigma}} \chi_{ik}^2 \sigma_j^2$,
where $\sigma_j$ are bosonic fields that only interact with $\chi_{ik}$. In
our notation, these $\sigma_j$ fields are nondissipative heat bath
fields\footnote{Such secondary interactions with $\phi$
also induce dissipative effects, though we will not treat that
here}.  Such interactions modify the
$\chi_{ik}$ decay width as
\begin{equation}
\Gamma_{\chi_{ik}}(T) \stackrel{{\rm add} \  \sigma  \  {\rm interactions}}
{\rightarrow}
\frac{T}{192 \pi} (g^4 +\frac{f^2}{8} + N_{\sigma} g^4_{\sigma}).
\end{equation}
If all the couplings $g^2_{\sigma} > 0$, then one must restrict the
number of $\sigma$ fields, $N_{\sigma}$, since their interaction with
$\chi_{ik}$ also adds a thermal mass contribution
$\approx g^2_{\sigma} N_{\sigma} T^2/12$.  In order to keep
$m_{\chi_{ik}} < T$, it requires 
$g^2_{\sigma} N_{\sigma} \stackrel{<}{\sim} 12$.
In the strong coupling limit $g^2_{\sigma} \sim 1$, this modification
with its stated limits increases $\Gamma_{\chi}(M)$ by a factor $10^{2}$.
In this case the bound on the Hubble parameter is
$H_{BI} \leq (2 \times 10^{-2}) \alpha \beta \kappa_M M$.

To go one step further, the thermodynamics of warm inflation 
in the previous sections would be unmodified if in total as many as
$N_{\sigma} \stackrel{<}{\sim} \beta \kappa_M N$ $\sigma$-fields coupled to
all the $\chi_{ik}$ fields that are thermally excited at a given
instant. In this case 
$H_{BI} \leq (2 \times 10^{-3}) \alpha \beta^2 \kappa_M^2 N M$.
{}For this case, precaution is necessary to keep
$m_{\chi_{ik}} < T$. One way to achieve this is if the 
$\chi_i-\sigma_j$ couplings are not sign definite.
Since the one-loop thermal mass correction is sensitive to
the sign $ \propto \pm g^2_{\sigma}$ whereas 
$\Gamma_{\chi} \propto g^4_{\sigma}$ is not, in principle
$\Gamma_{\chi}$ can be made arbitrarily large while the growth
of the thermal mass correction is controlled.
These modifications to $\Gamma_{\chi}$ may be useful for
phenomenological applications.

The final point addressed in this discussion is the general behavior in
all four cases of the density perturbation formulas.  The equation
of motion for mode $\delta \varphi_{{\bf k}_c}({\bf k}_p,\tau)$
can be obtained from a linearized approximation to
Eq. (\ref{eomgen}) and by replacing 
$\lambda \varphi_0^2 \rightarrow {\bf k}^2_p$ which gives
\begin{equation}
\frac{d \delta \varphi({\bf k}_p,\tau)}{d \tau}
\sim \frac{{\bf k}_p^2}{\eta'\Gamma' N g^2 \kappa_M^2}
D^{\gamma_{\eta}}(\tau)
\delta \varphi({\bf k}_p,\tau) \ \ + \ \ {\rm noise}.
\label{dphiapp}
\end{equation}
The decay function $D(\tau) \equiv (\varphi_0(\tau)/\varphi_{BI})$
and all exponents $\gamma$ are the same as defined in
Subsect. \ref{subsect4C}. Once again, our focus is not on the
(slowly varying) time dependence, but we have included the correct 
$\tau$-dependence for completeness.  The quantity multiplying
$\delta \varphi({\bf k}_p,\tau)$ on the right-hand-side of
Eq. (\ref{dphiapp}) is the decay rate for the respective mode.
Thus the freeze-out momentum for the modes is
\begin{eqnarray}
k_F & \sim & 
\left( \frac{H(\tau)}{\Gamma_{\chi}(M)}\right)^{1/2} (\eta'\Gamma')^{1/2}
N^{1/2} g \kappa_M D^{(\gamma_{\eta}+\gamma_{\rho''})/2}(\tau) M
\nonumber\\
& \sim & (\eta' \Gamma')^{1/2} \alpha^{1/2} \beta^{1/2} \kappa_M^{3/2}
N^{1/2} g \kappa_M D^{(\gamma_{\eta}+\gamma_{\rho''})/2}(\tau) M.
\label{kfgen}
\end{eqnarray}
Using this expression and the general expressions for
$d\varphi_0(\tau)/d\tau$, Eq. (\ref{phialt})  and
$\varphi_{BI}/M$, Eq.(\ref{pomgen}), implies
\begin{eqnarray}
\frac{\delta \rho}{\rho} & \sim & (\eta'\Gamma')^{3/4}
\frac{(H_{BI}/\Gamma_{\chi}(M))^{3/4}N^{3/4} g^{3/2}}{(N_r+2\kappa_MN)^{1/2}}
D^{(\gamma_{\eta}+\gamma_{\rho''}+2\gamma_T-4)/4}(\tau) (y\tau+1)
\nonumber\\
& \sim & (\eta'\Gamma')^{3/4}
\frac{\alpha^{3/4} \beta^{3/4} \kappa_M^{3/4} N^{3/4} g^{3/2}}
{(N_r+2\kappa_MN)^{1/2}}
D^{(\gamma_{\eta}+\gamma_{\rho''}+2\gamma_T-4)/4}(\tau) (y\tau+1).
\label{delrgen}
\end{eqnarray}

Some features are worth mentioning.  In Eq. (\ref{kfgen}), $k^2_F$ should be
proportional to the Hubble expansion rate $H_{BI}$, 
since a slower expansion rate allows longer relaxation time, thus modes of
lower $|{\bf k}_p|$ 
can equilibrate.  The dependence of $\delta \rho/\rho$ on $H_{BI}$
in Eq. (\ref{delrgen}) is less than the naive linear behavior given by the
defining formula Eq. (\ref{delrhof}). This arises because of $H_{BI}$ dependence
induced by the dynamics on the other factors, 
$k_F^{1/2} \propto H_{BI}^{1/4}$ and
$(T/\varphi_{BI})^{1/2} \propto H_{BI}^{-1/2}$.

\section{Examples}
\label{sect6}

In this section the solutions from the last two sections are
studied in a few examples. It is not the purpose of this paper to detail
the phenomenological consequences of these solutions.  However, here we
would like to obtain some idea about the absolute scales for the various
dimensional quantities and how they chance in various parametric
regimes as well as in the limit of increasing adiabaticity.
In much of this section, the numerical value of premultiplying constants
are evaluated in various expressions
and we set $\eta'=48$, $\Gamma'=1/192$.

Before turning to the examples, note that in our construction,
since the Hubble parameter, Eq.(\ref{hubdef}), is 
proportional to $\Gamma_{\chi}(M)$, the scales of all warm inflation quantities
are controlled by this decay rate.  Recalling our comments 
from the discussion Subsect. \ref{subsect5C} about
modifying ${\cal C}$ in $\Gamma_{\chi}(M)$,
in ``our model'' ${\cal C} = g^4 +f^2/8$ and in the ``extended model''
considered in Subsect. \ref{subsect5C} where $\sigma$-fields were
introduced ${\cal C}=g^4+f^2/8 + g^4_{\sigma}N_{\sigma}$.
In the three subsections that follow, we quote estimates for both
``our model'' and the ``extended model''.  All the estimates to follow
always are within the observationally consistent regime with respect to
expansion e-folds $N_e \stackrel{>}{\sim} 60$ and density perturbation
$\delta \rho / \rho \stackrel{<}{\sim} 10^{-5}$.

In Subsect. \ref{subsect6A} a general estimate of the
scales $M,T_{BI}, H_{BI}, m_{\phi_{BI}}$, and $i_{min}M$ are given
that encompasses the four cases studied in the last two sections. 
Subsects. \ref{subsect6B} and \ref{subsect6C} focuses on the case
studied in Subsects. \ref{subsubsect4B2} and \ref{subsubsect5B2}
for $V_0 =0$ and $T_{EI} \stackrel{<}{\sim} T_{BI}$.
In Subsect. \ref{subsect6B}, the dependence of the scales and
parameters are
examined in the limit of arbitrary adiabaticity $\alpha \rightarrow 0$.
{}Finally in Subsect \ref{subsect6C} the warm inflation solutions are
examined as the inflaton self-coupling parameter $\lambda$ varies over a
wide range including $\lambda \sim 1$.
Within the limits of the present analysis, 
we will find that $\lambda$ is restricted to
be tiny.  However, a possibility is examined that could increase
$\lambda$ up to $10^{-4}$, within a regime that is consistent with
observational requirements on e-folds and density perturbations.

\subsection{Estimate of Scale}
\label{subsect6A}

The absolute scale of all dimensional warm inflation quantities in our
solution are determined once the Planck mass is introduced
$M_p  \approx 10^{19} {\rm GeV}$.  Our solutions are constructed such
that all the dimensional quantities have been
expressed in terms of M.  To
determine $M$, note that the Hubble parameter $H_{BI}$ can be expressed
in two way, by its definition 
$H_{BI} \equiv \sqrt{8\pi \rho(0)/(3m_p^2)} \approx
\sqrt{8\pi \rho_v(0)/(3m_p^2)}$ and the expression 
Eq. (\ref{hubdef}) $H_{BI} = \alpha \beta \kappa_M \Gamma_{\chi}(M)$.
{}For $\rho_v(0)$, from the expressions for the ratio
$\rho_r/\rho_v$ Eqs. (\ref{rhorvA1}),(\ref{rhorvA2}),
(\ref{rhorvB1}), and (\ref{rhorvB2}), note that
$\rho_v(0) \approx N_e \rho_r(0) =N_e \pi^2 (N_r+2\kappa_MN)T^4_{BI}/16
= N_e \pi^2 (N_r+2\kappa_MN)\kappa_M^4 M^4/16$.
Using this in the first expression for $H_{BI}$ and Eq. ({\ref{chidecay})
in the second and equating the two, we find
\begin{equation}
M= \frac{\sqrt{3} \Gamma'}{\pi^{5/2}} \frac{\alpha \beta {\cal C}}
{\sqrt{N_e N (1+N_r/(2\kappa_MN))} \kappa_M^{3/2}} M_p.
\label{meqn}
\end{equation}
An approximate upper  bound can be obtained using 
Eqs. (\ref{thermaA1}), (\ref{thermaA2}), (\ref{thermaB1}),
and (\ref{thermaB2}) which
imply $\alpha \stackrel{<}{\sim} \beta/[200(1+N_r/(2\kappa_MN))]$
and Eqs. (\ref{forcecA1}), (\ref{forcecA2}),
(\ref{forcecB1}), and (\ref{forcecB2}) which imply
$\kappa_M \stackrel{>}{\sim} 10\sqrt{1+N_r/(2\kappa_MN)}/\beta$.
Substituting these bounds into the above equation (\ref{meqn}) gives
\begin{equation}
M \stackrel{<}{\sim}  \frac{(8.2 \times 10^{-8}) \beta^{7/2} {\cal C}}
{\sqrt{N_e N} (1+N_r/(2\kappa_MN))^{9/4}} M_p.
\label{minequal}
\end{equation}
{}From this expression and the bound on $\kappa_M$ it follows that
\begin{equation}
T_{BI} \stackrel{<}{\sim} \frac{(8.2 \times 10^{-7}) \beta^{5/2} {\cal C}}
{\sqrt{N_e N} (1+N_r/(2\kappa_MN))^{7/4}} M_p,
\label{tinequal}
\end{equation}
\begin{equation}
H_{BI} \stackrel{<}{\sim} \frac{(6.8 \times 10^{-12}) \beta^{9/2} {\cal C}^2}
{\sqrt{N_e N} (1+N_r/(2\kappa_MN))^{11/4}} M_p,
\label{hubinequal}
\end{equation}
\begin{equation}
m_{\phi_{BI}} \stackrel{<}{\sim} 
\frac{(4.4 \times 10^{-8}) \beta^{5/2} (1-\beta)^{1/2} g {\cal C}}
{N_e (1+N_r/(2\kappa_MN))^{4}} M_p,
\label{mpinequal}
\end{equation}
and $g \varphi_{EI}$ or equivalently the scale of the mass levels
$\sim i_{\rm min}  M$ in the DM-model is
\begin{equation}
i_{\rm min} M  \stackrel{<}{\sim} 
\frac{(1.0 \times 10^{-4}) \beta^{3 \pm {\rm few}} N_e^{1/2} {\cal C}}
{(1-\beta^{O(1)}) N^{1/2} (1+N_r/(2\kappa_MN))^{3/4}} M_p.
\label{imininequal}
\end{equation}

To quote some numbers, consider a typical case with $\beta =0.5$,
$N_e=60$, $N_r=0$, $N=5$ and ``our model'' with $f \approx 1$ so that
${\cal C} \approx 1/8$ (``extended model'' with $g_{\sigma} \approx 1$,
$N_{\sigma} \approx 12$, so that ${\cal C} \approx 12$).
{}For this case we find $M \stackrel{<}{\sim} 5.2 \times 10^8 {\rm GeV}$
($\stackrel{<}{\sim} 5.0 \times 10^{10} {\rm GeV}$),
$H_{BI} \stackrel{<}{\sim} 2.7 \times 10^3 {\rm GeV}$
($\stackrel{<}{\sim} 2.5 \times 10^7 {\rm GeV}$),
$T_{BI} \stackrel{<}{\sim} 1.0 \times 10^{10} {\rm GeV}$
($\stackrel{<}{\sim} 1.0 \times 10^{12} {\rm GeV}$),
$i_{\rm min} M \stackrel{<}{\sim} 1.2 \times 10^{14} {\rm GeV}$
($\stackrel{<}{\sim} 1.0 \times 10^{16} {\rm GeV}$),
and $T_{EI} =T_{BI}/2$.  We also 
find $N_M \stackrel{>}{\sim} 2.6 \times 10^5$ mass
sites are crossed.  If we set $\kappa_M$ at its lower bound,
$\kappa_M = 1/\sqrt{2\alpha \beta}$, and require
$\delta \rho/\rho \leq 10^{-5}$, then in all four cases within
the high temperature regime, $k_F < T$, it requires
$g \leq 0.2$. {}From this it follows that $\lambda \leq 10^{-16}$ and
$m_{\phi_{BI}} \stackrel{<}{\sim} (2.1 \times 10^{-3}) T_{BI}$.
The high temperature validity regime for the
density perturbation results in Sect. \ref{sect5} require $k_F < T$.
{}From the expressions for $k_F$ in the four cases, we find in general
$k_F \sim \sqrt{N_e} m_{\phi}$.  Thus the estimates given here involving
density perturbations are valid for $N_e < 2 \times 10^5$.
{}Finally, for both our and the extended models, the thermalization rate
$\Gamma_{\chi}(T)$ is about 400 times faster than the Hubble expansion
rate $H(\tau) \stackrel{<}{\sim} H_{BI}$, 
so that the thermalization approximation is
well satisfied.

Although $\lambda$ is tiny, the inflaton mass, $m_{\phi}$ is
large relative to the Hubble parameter.  In the above example $m_{\phi}$
is three orders of magnitude below the temperature scale but four orders
of magnitude larger than the Hubble parameter.  The smallness of $\lambda$
preempts questions about fine tuned potentials, similar to the situation
in supercooled dynamics. This point briefly is addressed in
Subsect. \ref{subsect6C}.  However, it should be noted that for these
tiny values of $\lambda$, when the thermal damping is removed after
the mass site $M_{i_{\rm min}}$, the potential does not support
inflation. Once the thermal damping is removed, the only damping term
that remains is due to the coupling of the inflaton to the background
cosmology. This yields a $3H {\dot \varphi}_0$ term that is familiar
from supercooled inflationary dynamics.  The inflaton equation of motion
then becomes $3Hd\varphi_0/dt = -\lambda \varphi_0^3/6=
-m_{\phi}^2 \varphi_0/3$. Thus in a Hubble time $\Delta t = 1/H$,
$|\Delta \varphi_0/\varphi_0| \approx m_{\phi}^2/(9H^2) \gg 1$, so that
$\varphi_0$ rapidly falls down the potential.  In words, the curvature
of the potential is huge relative to the scale of the Hubble expansion
rate.  As such, to terminate warm inflation in this model and go into a
radiation dominated regime, it suffices simply to stop coupling the
inflaton to mass sites.

\subsection{The Limit of Arbitrary Adiabaticity}
\label{subsect6B}

The self consistency of the near-thermal-equilibrium quantum field
theory formalism applied in this paper requires satisfying the conditions in
Subsect. \ref{subsect3B}.  Based on the solutions in the
previous two sections, we find that the most constraining consistency
conditions are the thermalization and thermal-adiabatic conditions
Eqs. (\ref{therm}) and Eq. (\ref{thermadiab}) respectively.  As it turns out, both these
conditions are controlled by a single parameter in our solutions,
$\alpha$, with the validity for both conditions improving as
$\alpha \rightarrow 0$.  In this subsection the observationally
consistent regime with respect to $N_e$ and $\delta \rho/\rho$ is
studied as a function of $\alpha$, in particular, in the limit of
arbitrary adiabaticity $\alpha \rightarrow 0$.  We focus on the
case in Subsects. \ref{subsubsect4B2} and \ref{subsubsect5B2}
for $V_0=0$, $T_{EI} \stackrel{<}{\sim} T_{BI}$.

{}For the case of interest, from
Subsect. \ref{subsubsect4B2} the consistency conditions are
\begin{equation}
\alpha < \frac{1}{2\beta \pi^4}
\end{equation}
and
\begin{equation}
\kappa_M > \frac{1}{2\sqrt{2}\alpha^{1/2} \beta^{1/2}}, 
\label{kaplimB2}
\end{equation}
where throughout this subsection we set $N_r=0$.
The mass splitting scale parameter M is determined by the same
procedure as in the previous subsection.  We find 
\begin{equation}
M=\frac{3\Gamma'}{2\pi^{5/2}} 
\frac{\alpha(1-\beta)^2 \beta^{1/2} {\cal C}}
{\sqrt{NN_e}\kappa_M^{3/2}} M_p.
\label{m6b}
\end{equation}
The other dimensional quantities can be determined easily from
this.  

In Fig. 1, the $\alpha$ dependence of all dimensional scales are shown
for two cases.
$\kappa_M$ is set to its lower
bound, $\kappa_M = 1/(2\sqrt{2}\alpha^{1/2} \beta^{1/2})$, with always the
restriction
$\kappa_M>1$.  
The limit of arbitrarily increasing adiabaticity is $\alpha \rightarrow
0$ ($-\log_{10}(\alpha) \rightarrow \infty$).  All the scales,
$M,T_{BI},H_{BI},i_{\rm min}M$, and $m_{\phi_{BI}}$, are in GeV with
their $\log_{10}$ plotted.  The solid lines are for the case
$N=5$, $N_e=65$, $\beta=0.5$
and ${\cal C}=1/8$ (``our model'').
The dashed lines are for the case
$Ng^4=1/8$, $N_e=65$, $\beta=0.5$
and ${\cal C}=1/8$ (``our model'').
{}For the ``extended model'',
${\cal C}=12$, all scales in Fig.1 are shifted up by a 
factor $\sim 10^2$ except for $H_{BI}$ which is shifted up by a factor
$\sim 10^4$.
For the region to the left of the dotted vertical line at
0.6 ($\alpha > 0.25$), $\kappa_M=1$.

$M,T_{BI},H_{BI}$, and $i_{\rm min}M$ are independent of $\delta
\rho/\rho$, whereas $m_{\phi} \propto g$, so it depends on
$\delta \rho/\rho$ since from Eq. (\ref{delrB2})
\begin{equation}
g \approx 30.45 \frac{(\delta \rho/\rho)^{2/3}}
{\alpha^{5/12} \beta^{5/12} N^{1/6}}.
\label{g6b}
\end{equation}
For $\alpha \rightarrow 0$ with everything else fixed, $g$ increases.
Since $g<1$ is required by perturbation
theory, the model requires $\delta \rho/\rho \rightarrow 0$ as
$\alpha \rightarrow 0$.
In Fig. 1, we set 
$\delta \rho/\rho= 10^{-5}$ down to the smallest $\alpha$ possible,
which is given by the vertical solid and dashed lines for
the two respective cases.  
To the right of these lines (smaller $\alpha$), 
$\delta \rho/\rho$ is less
than $10^{-5}$.
{}For the other parameters in the model, as $\alpha$ ranges from 0 to 1
($-\log_{10}(\alpha)$ ranges from $\infty$ to 0), for the
solid case the ranges are $g$ from 1 to 0.013, $\lambda$ from 
0 ($\propto \alpha$) to $4 \times 10^{-17}$, $\kappa_M$ from
$\infty$ ($\propto 1/\alpha^{1/2}$) to 1, and $N_M$ from $\infty$
($\propto 1/\alpha$) to 350; for the dashed case the ranges are
$g$ from 0.59 to $1 \times 10^{-5}$, $N$ from
1 to $3 \times 10^{18}$, $\lambda$ from 
0 ($\propto \alpha$) to $4 \times 10^{-11}$, $\kappa_M$ from
$\infty$ ($\propto 1/\alpha^{1/2}$) to 1, and $N_M$ from $\infty$
($\propto 1/\alpha$) to 350.

{}For $\alpha < 1/\pi^4$, which is
in the region to the right of the dot-dashed vertical
line at $-\log_{10}(\alpha)=1.9$, all 
adiabaticity conditions are valid.  To the left of this
vertical line, the thermal-adiabatic condition in its stringent form
is not valid.  However, as discussed in Subsect. \ref{subsect3C},
the thermal-adiabatic condition may still hold in some part
of this region.  To determine the extent to which the thermal-adiabatic
condition can be relaxed requires details about
thermalization that go beyond the simple high-temperature
approximations applied in this paper.

The DM-model warm inflation calculation in \cite{bgr2} is similar to the
case in Subsect. \ref{subsubsect4B2} for $V_0=0$, 
$T_{EI} \stackrel{<}{\sim} T_{BI}$.
The difference is \cite{bgr2} ignores the
minor modifications that arise due to
time dependence of the
temperature, whereas this was treated in Subsect. \ref{subsubsect4B2}.
The region studied in \cite{bgr2} is for $0.5 < \alpha \leq 1$.
This was chosen for its simplicity in illustrating
the basic features of the results.
In this region, all the consistency conditions are satisfied except
the stringent form of the thermal-adiabatic condition.  Based on earlier
discussions in this paper, this region is still within the plausible
validity region.  We have verified that in the region of overlap 
the results in \cite{bgr2} agree with thoses in
Subsect. \ref{subsubsect4B2}.

\subsection{$\lambda$ Dependence of the Solution}
\label{subsect6C}

In this subsection, we examine the $\lambda$ dependence of our
solution for the case from Subsects. \ref{subsubsect4B2} and
\ref{subsubsect5B2}.  {}For this, we treat $\lambda$ as an independent
variable in exchange for $\kappa_M$, which from Eq. (\ref{lamB2}) gives
\begin{equation}
\kappa_M = (1.2 \times 10^8)
\frac{\beta^3 N_e^3 \lambda}{(1-\beta)^3 N g^4}
\label{kappalam}
\end{equation}
where $\alpha = (2\beta \pi^4)^{-1}$ has been set to its upper bound
and we only consider the regime with $N_r=0$.
Here and throughout this subsection, the numerical values of all
constants are quoted and we have used $\Gamma'=1/192$, $\eta'=48$.
The parameters on the right-hand-side of Eq. (\ref{kappalam})
can be varied freely up to the mild constraints on $\kappa_M$,
Eq. (\ref{kaplimB2}).  Substituting Eq. (\ref{kappalam}) into our solutions
in Subsects. (\ref{subsubsect4B2}) and (\ref{subsubsect5B2})
we find
\begin{equation}
M=(1.8 \times 10^{-18}) \frac{(1-\beta)^{13/2} Ng^6 {\cal C}}
{\beta^5 N_e^5 \lambda^{3/2}} M_p,
\label{mlam}
\end{equation}
\begin{equation}
T_{BI}=(2.1 \times 10^{-10}) \frac{(1-\beta)^{7/2} g^2 {\cal C}}
{\beta^2 N_e^2 \lambda^{1/2}} M_p,
\label{tlam}
\end{equation}
\begin{equation}
H_{BI}=(1.8 \times 10^{-15}) \frac{(1-\beta)^{7/2} g^2 {\cal C}^2}
{\beta^2 N_e^2 \lambda^{1/2}} M_p,
\label{hlam}
\end{equation}
\begin{equation}
m_{\phi_{BI}}=(3.9 \times 10^{-8}) \frac{(1-\beta)^{5/2} g {\cal C}}
{\beta N_e } M_p,
\label{mplam}
\end{equation}
and $N_M=(3 \times 10^{10}) \beta^4 (1-\beta^{3/4}) \lambda N_e^4/[
(1-\beta)^4 N g^4]$ so that
\begin{equation}
i_{\rm min} M=(5.5 \times 10^{-8}) 
\frac{(1-\beta)^{5/2} (1-\beta^{3/4}) g^2 {\cal C}}
{\beta^{17/4} N_e \lambda^{1/2}} M_p.
\label{imlam}
\end{equation}

Also we express $g$ in terms of $\lambda$ and $\delta \rho/\rho$ to
obtain
\begin{equation}
g= (4.4 \times 10^3) \frac{(1-\beta)^{3/2} (\delta \rho/\rho)^2}
{\beta^{3/2} N_e^{3/2} \lambda^{1/2} }
\label{gdelr}
\end{equation}
This can be substituted in the above expressions for the scales,
but the perturbative restriction must be respected $g<1$.
In the observationally consistent regime, 
$\delta \rho/\rho \stackrel{<}{\sim} 10^{-5}$, this perturbative restriction is
comfortably satisfied for a wide range of the parameters.

It is interesting to examine the maximum size of $\lambda$ within the
observationally consistent regime of warm inflation.  {}For satisfying
just the horizon/flatness problems, the $\lambda \sim 1$ regime has
solutions if also $g \sim 1$.  However the primary restrictions arise
from the density perturbation constraints.   If we restrict to only the
high temperature regime for the freeze-out momentum,
the inequality  $k_F < T$ must be imposed to Eq. (\ref{kfB2}).
Reordering this inequality to isolate $\lambda$ and using 
the solutions from Subsect. \ref{subsubsect4B2}, we find the constraint
\begin{equation}
\lambda < (6.6 \times 10^{-5}) \frac{(1-\beta)^3 g^2}{\beta^3 N_e^3}.
\label{lamlim}
\end{equation}
Setting $\lambda$ at its upper limit and substituting into Eq. (\ref{gdelr}),
for $N_e=65$, $\delta \rho/\rho = 10^{-5}$, $\beta=0.5$,
we find $g \approx 7.2 \times 10^{-3}$, which from
Eq. (\ref{lamlim}) implies $\lambda \approx 1 \times 10^{-14}$.
This value of $\lambda$ is two orders of magnitude larger than the limit
in Subsect. \ref{subsect6A}, because here 
$\kappa_M \approx (1.4 \times 10^8)/N$ is not at its lower bound.
With these values for $\lambda$,$g$, $N_e$, $\beta$ , and leaving $N$
unspecified,
from Eqs.
(\ref{mlam}) - (\ref{imlam}) for ''our model'' with ${\cal C}=1/8$
(''extended model'' with ${\cal C}=10$)  we find the scales in GeV
$M \approx 0.7N (6N)$, $T_{BI} \approx 1 \times 10^7$ ($8 \times 10^8$), 
$H_{BI} \approx 11$ ($7 \times 10^4$),
$m_{\phi_{BI}} \approx 2 \times 10^6$ ($2 \times 10^8$),
and $N_M= 10^{12}/N$ so that
$i_{\rm min} M \approx 2 \times 10^{11}$ ($5 \times 10^{13}$).

{}For $k_F > T$, as stated earlier, the high-temperature calculations in
the previous section are not valid.  Suppose that thermal dissipation is
inactive for wavenumbers larger than the temperature $k_F > T$.
In this case, if the freeze-out wavenumber from Eq. (\ref{kfB2}) is
larger than $T$, the $\varphi$-amplitude should be set to its limiting
value $\Delta \varphi^2_H \approx T^2/(2\pi^2)$. In this case, the 
density perturbation expression in Subsect. \ref{subsubsect5B2}
is replaced by 
$\delta \rho/\rho \approx 3 \alpha^{1/2} \beta^{1/2} g (y\tau+1)^{3/10}/
(5\pi^3)$.  {}From the previous paragraph, $k_F > T$ corresponds to
the region of $\lambda$ in Eq. (\ref{lamlim}) with $<$ replaced by
$>$.

{}For this hypothetical case, if we set $N_e=65$ and $\beta=0.5$, we find
$\kappa_M \approx (1 \times 10^{22}) \lambda /N$,
$M \approx (8 \times 10^{-41}) N {\cal C} M_p/\lambda^{3/2}$,
$T_{BI} \approx (9 \times 10^{-19}) {\cal C} M_p/\lambda^{1/2}$,
$H_{BI} \approx (8 \times 10^{-24}) {\cal C}^2 M_p/\lambda^{1/2}$,
$m_{\phi_{BI}} \approx (2 \times 10^{-12}) {\cal C} M_p$,
and $N_M \approx (8 \times 10^{25}) \lambda/N$ so that
$i_{\rm min}M \approx (6 \times 10^{-14}) {\cal C} M_p/\lambda^{1/2}$.
If the temperature scale for inflation is assumed to be above 1 TeV,
it requires ${\cal C}/\lambda^{1/2} > 100$, which for ${\cal C}=10$
implies the self-coupling can be as large as $\lambda \sim 0.0001$, 
although it also
requires a very large number of fields.  In any case, it is interesting
to examine the difficulties that must be overcome in this model to avoid
an ultra-flat inflationary potential.  The fact that such a possibility
in remotely realizable is interesting and motivates further
investigation.

In summary, the magnitude of $\lambda$ is dependent mutually on the
overdamped dynamics of warm inflation and the density perturbation
requirement.  {}Further development of the simple dynamical framework used here
may increase $\lambda$ by several orders.

\section{Conclusion}
\label{sect7}

In this paper, warm inflation solutions have been obtained for the DM
model that solve the cosmological horizon, flatness and scalar density
perturbation problems.  {}Furthermore such solution regimes exist for an
arbitrarily slow evolution of all macroscopic variables in the inflaton
field system and the background cosmology.  {}For convenience, the DM
models considered in this paper had all adjacent mass sites equally
spaced, $g|M_{i+1}-M_i| = M$.  This condition is not required to obtain
warm inflation solutions.  Spacings between adjacent sites can widely
vary.  The inflaton motion remains overdamped for a succession of sites
in which adjacent spacings are all less than the temperature $T$.
Spacings larger than $T$ are not inconceivable, but the inflaton motion
is more complicated.  {}For any DM model spacings, the ultimate test is
the the usefulness of warm inflation solutions that it yields.  We
have examined only the single case of equal spacings.

There are a few improvements to these calculations
that can be made.  Our calculations have adhered to the high-temperature
approximation, which means fields with mass $m \leq T$ are thermally
active and those with $m> T$ are thermally dormant. This approximation
is over restrictive.  Generally, fields participate in
dissipative and thermalization dynamics once $m \leq 10T$.
An example that treated dissipative heat bath fields with $m \leq 2.5T$ is
in \cite{bgr2}.  It is worthwhile to extend the present
calculation beyond the high-temperature approximation.
Another improvement to this calculation is to compute the ladder
resummed dissipative function similar to the shear viscosity case 
computed by Jeon in
\cite{jeon}. This point was noted in \cite{bgr} and recently
has been verified by Jeon \cite{jeonpc}.

The model studied in this paper has been developed into a string theory
warm inflation scenario in \cite{bk,bk2}.  The DM model has an essential
feature for this interpretation.  Its hierarchy of mass sites are
reminiscent of the tower of mass levels of a string. 

The first step towards this interpretation was to obtain the DM model
from a SUSY superpotential \cite{bk}.  This has a relevance independent
of the string interpretation.  It establishes that the DM model is
natural in the sense of nonrenormalization theorems, which means once
the parameters are chosen, they stay fixed until SUSY is broken.

The string picture developed in \cite{bk,bk2} interprets the DM model as
an effective SUSY model in which the inflaton is a string zero mode and
it interacts with higher string mass levels, which are the dissipative
heat bath fields.  Since the multiplicity of degenerate string
states increases
exponentially with excitation level, strings can provide an adequate supply
of dissipative heat bath fields.  Interestingly, the dispartity in
scales that generally arises in DM model warm inflation realizations,
$i_{\rm min} M \gg T_{BI} > M$, is readily explained in the 
string interpretation.  $i_{\rm min} M$ of the DM model corresponds
in the string interpretation to the mass scale of a string level for the
unperturbed string, i.e., when the coupling to $\phi$ is switched off.
The mass splitting scale $M$ corresponds in the string interpretation to
a fine structure splitting of a initially degenerate string mass level.
This fine structure splitting of the level arises from symmetry breaking.
As temperature drops below the scale of the string mass levels,
generally degeneracies in the mass levels will be lifted and
thereby create fine structure splittings.
Since for the warm inflation solutions in this paper 
$T_{BI} \ll i_{\rm min} M$, the conditions are adequate for
various perturbations to break the degeneracy of the mass level.

{}For example, if a symmetry breaking 
occurs at scale $i_{\rm min} M > v_1 > T_{BI}$,
the states of any mass level split by
characteristic scale $v_1$. Thus after symmetry
breaking, the states of an initially degenerate mass level shift
within a width of order $v_1$. A generic symmetry breaking typically will not
lift all the degeneracy, so that a mass level with ${\cal N}$
degenerate states before symmetry breaking splits into 
${\cal D} < {\cal N}$ finely split levels.  In this case,
the fine structure
splitting scale is $M \sim v_1/{\cal D}$.  Examples of symmetry
breaking scenarios and estimates of ${\cal N}$, ${\cal D}$ and $v_i$
are given in \cite{bk2}.

{}For the standard type I, II, heterotic and bosonic strings, the string
scale is $M_S \sim 10^{17} {\rm GeV}$. Therefore the string
interpretation requires $i_{\rm min} M \sim 10^{17} {\rm GeV}$.
{}For the first mass level, $n=1$, of the heterotic string,
for example,  ${\cal N}\approx
2 \times 10^{7}$ states.  ${\cal D}$ depends on the specific
symmetry breakings  that occurs.  {}For a GUT motivated example in
\cite{bk2} we found ${\cal D} \stackrel{<}{\sim} 10^5$.

Comparing these estimates to the bounds in Subsect. \ref{subsect6A},
$i_{\rm min}M$ is at least one order of magnitude below $M_S$.  
The example in \cite{bk2} found $v_1/M \approx 10^5$. Since 
$T_{BI} < v_1$, to satisfy the example 
in \cite{bk2} requires $T_{BI}/M < 10^5$.
In Subsect. \ref{subsect6A} we found $T_{BI}/M \approx 10^2$ with
$T_{BI} \approx 10^{10-12} {\rm GeV}$ so that $v_1 \approx 10^{13-15}
{\rm GeV}$.  Thus the estimates from the model in this paper are
somewhat compatible with the string scenario in \cite{bk2}.  $i_{\rm
min}M$ in our model still is below the string scale $M_S$. $v_1$ is
within range of the GUT scales.  The improvements to this
calculation discussed above should elevate $i_{\rm min}M$ to $M_S$ and
allow greater flexibility in the range of $T_{BI}$ and $M$.

This good news comes at the expense of a finely tuned self-coupling
parameter $\lambda$. The nature of this fine tuning problem is different
from a similar problem in supercooled inflation scenarios
\cite{ni,ci}.
In our model, the dynamical parameter $m_{\phi}^2 \propto \lambda
\varphi_0^2$ is quite large relative to 
the scale of the Hubble parameter. {}From
Subsect. \ref{subsect6A} we find in our model $m_{\phi}/H_{BI} \approx
10^{2-4}$.  In contrast, in supercooled scenarios 
$m_{\phi} \stackrel{<}{\sim} H$.  {}Furthermore Subsect. \ref{subsect6C}
demonstrated that $\lambda$ need not be tiny for observationally
interesting warm inflation.  However this required all the scales to be
much smaller and it required an extremely large number of fields.  Both
these requirements conflict with the properties of conventional strings.
These are disappointing features of this model.  Nevertheless, there are
indications that in this warm inflation dynamics, the density
perturbation requirements do not mandate a tiny $\lambda$.
Dissipative dynamics provides other means for bounding the density
perturbation amplitude.  The present model, despite some of the
shortcomings, demonstrates the importance of the full decohereing
dynamics in determining the density perturbations.

In this paper, particle production has been treated to the minimal extent
of its implications for strong dissipation and solving the cosmological
puzzles.  This comes short of attempting a detailed modeling of the
produced particle spectrum.  This direction 
has some interesting possibilities,
which will be discussed here. 

An important fact to note is that
the heat bath particles produced in this scenario are generally very massive,
$ \sim 10^{16} - 10^5 {\rm GeV}$.
Recall in this model, as the inflaton
amplitude $\varphi_0$ relaxes to its minimum,
the heat bath fields at a given site $i$ become thermally excited
when their mass Eqs. (\ref{rmchi}), (\ref{rmpsi})
$m_{\chi_{ik},\psi_{ik}} \propto |\varphi_0-M_i|$ 
falls below the
temperature and otherwise they are thermally unexcited.
In order to realize this picture, an elementary requirement
is that the microscopic time scale of thermalization
must be faster than the characteristic time scale for the mass of the
particle to change, as reflected through the thermal-adiabatic
condition Eq. (\ref{thermadiab}). In general this condition becomes
increasingly less valid for a given heat bath field, as its
mass becomes increasingly larger than the temperature,
because its decay width $\Gamma_{\chi,\psi}$ decreases
exponentially $\propto \exp(-m_{\chi,\psi}/T)$.
Thus, above some $m_{\chi,\psi} > T$ the thermal-adiabatic condition
no longer will be satisfied.  At this point, the existing abundance
of the given heat bath particles then will freeze-out.  After
freeze-out, the mass of these heat bath particles
continues to grow, due to their $\varphi_0$ dependence,
but their total number is fixed.  Since prior to freeze-out the
thermal distribution of the heat bath fields also falls
exponentially $\propto \exp(-m_{\chi,\psi}/T)$,
at time of freeze-out, their abundance could be large
if $m_{\chi,\psi} \sim T$ or small if
$m_{\chi,\psi} \stackrel{>}{\sim} 10T$.  As such,
the magnitude of the freeze-out abundance depends on
how quickly the thermal adiabatic condition becomes
invalid, and our solutions cover a wide range of possibilities.
The main point to observe here is that this dynamics generically
produces very massive particles, in the range
$ \sim 10^{16} - 10^5 {\rm GeV}$ once $\varphi_0$ equilibriates.
Such heavy particle production at the early stages of warm
inflation is not important in the post-inflation universe,
since they will completely dilute during the inflationary
period.  However, this heavy particle productions from
the last few mass sites, just before warm inflation ends,
will leave significant abundances in the post-inflation
universe.    

Thus, without further modifications to 
this model, particle production yields weakly 
interacting massive particles (WIMPs), which are a category 
of particles that play an important role
in present-day notions about dark matter \cite{wimps}.
In fact, exceptionally high mass particles 
$\sim 10^{12} - 10^{16} {\rm GeV}$, which
corresponds to the upper mass range for the
heat bath fields, sometimes
are further distinguished
in the literature as ``Wimpzillas'' \cite{wimpzilla}. 

In general, thermal production of very massive WIMPs
is understood to lead to overabundance problems, which therefore
restrict the upper bound on their mass to be
$\stackrel{<}{\sim} 10^6 {\rm GeV}$ \cite{griest}.  
This implies thermal Wimpzilla production is prohibited
for any generic situation, including the model in
this paper.  Thus most of the mass range for
the heat bath fields in our model is in danger of
an overabundance problem.   
In fact, as mentioned above, for the model
in this paper the problem is worse since upon
production, the mass of the massive heat bath particles
can grow by as much as several orders of magnitude
due to the $\varphi_0$-dependence in Eqs. (\ref{rmchi}), (\ref{rmpsi}).

There are simple remedies to this overabundance problem, which provide
a rich number of phenomenological possibilities
in which the abundance of WIMPs and Wimpzillas can be controlled.
The simplest solution is to couple the heat bath fields to secondary
fields, such as the example of $\sigma$ fields in Subsect. VC.
This provides decay channels for the 
$\chi,\psi$ heat bath fields
into less massive or even light fields, thus providing a
means to control the relative fractions of
WIMPs, Wimpzillas, light particles etc... Furthermore, as discussed
in Subsect. VC, this mechanism can be implemented with minimal effect on
the dissipative dynamics.  Another possibility, which can be applied
separately or in conjunction with this,
is to have a second and short stage of inflation after the initial warm 
inflation stage, which dilutes the WIMP and Wimpzilla abundances.
Finally, the lower end of the  mass range of the heat bath fields
found from our calculations falls below
the upper acceptability limit for thermal WIMP production.
The present model has a narrow mass window that falls within
this acceptability limit.  However, with the improvements 
discussed earlier in this section, to the dissipative
dynamics, this mass range could
be substantially enlarged.

The calculations in this paper were guided by observation and
consistency with quantum field theory.  ``Nice'' particle physics was
not an {\it a priori} requirement.  On the one hand, this is a constructive
approach that attempts to find a toy model to study a very complicated
dynamics.  On the other hand, this is a predictive approach. Quantum
field theory is postulated to hold at the scales of inflationary
dynamics. Combining this theoretical tool with observation, a mutually
consistent model has been deduced.  

We can conclude that this model is a good
constructive tool for studying warm inflation dynamics.
No complete conclusion is possible as yet about the predictive content of this
model.  The interesting connection between this model and strings found
in \cite{bk,bk2} implies a revised meaning is required of ``nice''
particle physics.  At the inflationary 
scale, nice particle physics may not be synonymous with simple particle
physics.  This model is evidence that inflationary dynamics can be a
multi- or even ultramulti- field problem.

\section*{Acknowledgments}
I thank R. Brandenberger, M. Gleiser, A. Guth, S. Jeon, T. Kephart,
A. Linde, R. Ramos, M. Sher and J. Yokoyama for helpful discussions.
This work was supported by the U. S. Department of Energy
under grant DE-FG05-85ER40226.

\newpage

\begin{center}
FIGURE CAPTION
\end{center}

\vspace{0.5cm}

{\bf Figure 1}:  Warm inflation scales as a function of the adiabatic
parameter $\alpha$ for
solid lines:
$N=5$, $N_e=65$, $\beta=0.5$
and ${\cal C}=1/8$;
dashed lines:
$Ng^4=1/8$, $N_e=65$, $\beta=0.5$
and ${\cal C}=1/8$.  For the scalar density perturbations, 
$\delta \rho/\rho = 10^{-5} \ (< 10^{-5})$ to the left (right) of the vertical
solid and dashed lines for the two corresponding cases.
For the entire $\alpha$ range, the spectrum of density perturbations is flat
up to logarithmic corrections.

\begin{figure}[b]
\epsfysize=18cm 
{\centerline{\epsfbox{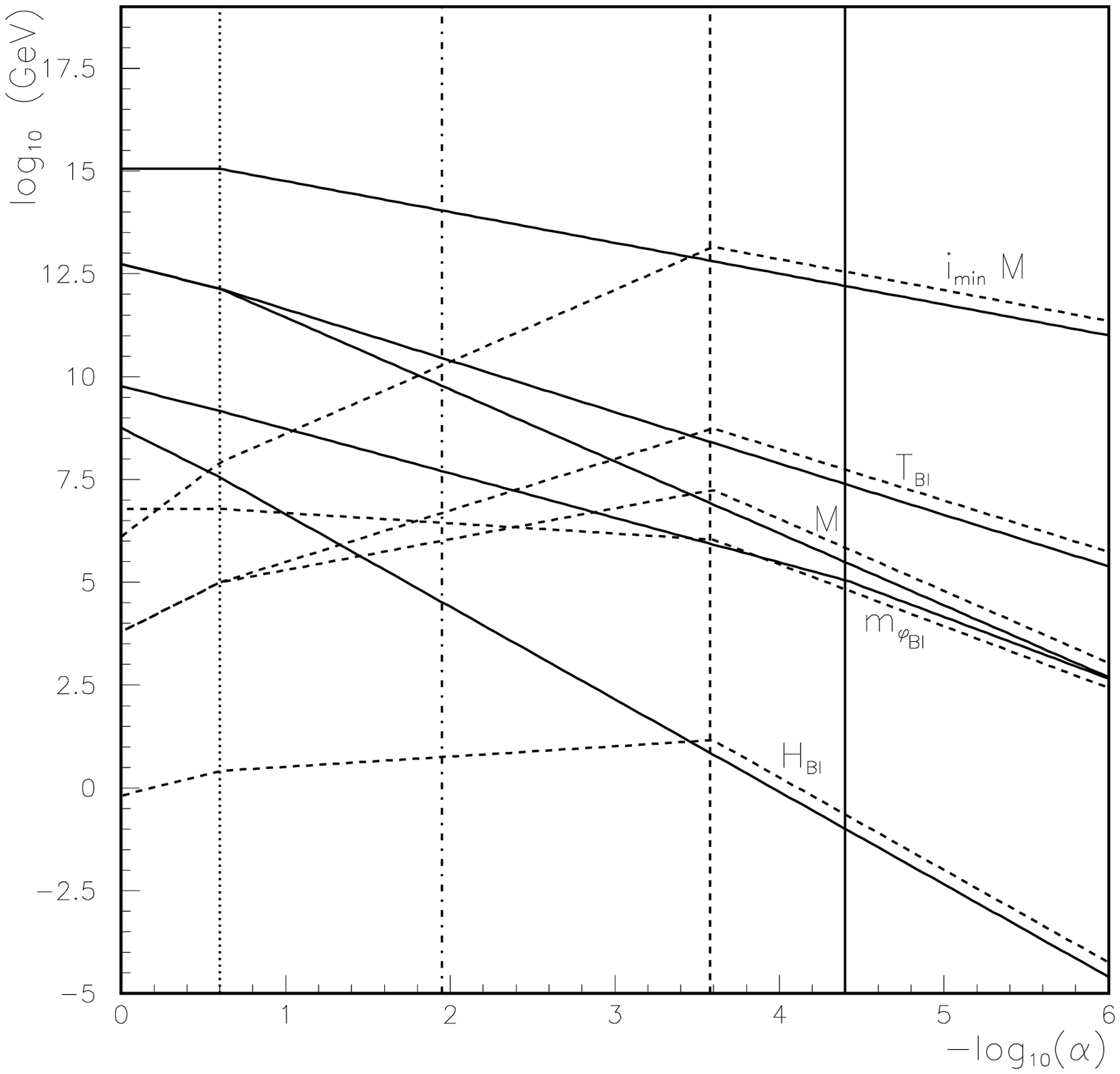}}}

\vspace{1cm}

\end{figure}

\newpage
\end{document}